\documentclass[reprint,amsmath,amssymb,aps,twocolumn,longbibliography]{revtex4-1}
\usepackage{graphicx}
\usepackage{dcolumn}
\usepackage{rotating}
\usepackage{amsthm,amsmath,amssymb,amsfonts,bbm}
\usepackage{mathptmx}
\usepackage{mathrsfs}
\usepackage{xcolor}
\usepackage{siunitx}
\usepackage{bm}
\usepackage{textcomp}
 
\def\omr{\omega_\textrm{r}}
\def\ommm{\omega_\textrm{mm}}
\def\omc{\omega_\textrm{c}}
\def\omm{\omega_\textrm{mod}}
\def\omp{\omega_\textrm{p}}
\def\omf{\omega_\textrm{f}}
\def\omg{\omega_\textrm{gap}}

\def\ke{\kappa_\textrm{ext}}

\def\ai{a_\textrm{in}}
\def\ao{a_\textrm{out}}

\def\vq{\textbf{q}}
\def\vk{\textbf{k}}
\def\vr{\textbf{r}}

\def\meana{\langle a\rangle}

\def\meansz{\langle \sigma_z\rangle}
\def\meansx{\langle \sigma_x\rangle}
\def\meansy{\langle \sigma_y\rangle}
\def\meann{\langle n \rangle}

\def\rla{\rangle\langle}
\def\me{m_\textrm{e}}

\def\tr{\textrm{tr}_\textrm{R}}
\def\Te{T_\textrm{e}}

\def\ns{n_\textrm{s}}

\def\kB{k_\textrm{B}}

\begin{document}

\title{Sensitive detection of the Rydberg transition in trapped electrons on liquid helium using radio-frequency reflectometry}

\author{Jui-Yin Lin} 
\altaffiliation[Current address: ]{Fluid Mechanics Unit, Okinawa Institute of Science and Technology (OIST) Graduate University, Onna, 904-0495 Okinawa, Japan}
\affiliation{Quantum Dynamics Unit, Okinawa Institute of Science and Technology (OIST) Graduate University, Onna, 904-0495 Okinawa, Japan}
\author{Tomoyuki Tani} 
\altaffiliation[Current address: ]{Department of Applied Physics, Hokkaido University, Sapporo, 060-8628 Hokkaido, Japan}
\author{Mikhail Belianchikov} 
\affiliation{Quantum Dynamics Unit, Okinawa Institute of Science and Technology (OIST) Graduate University, Onna, 904-0495 Okinawa, Japan}
\author{Denis Konstantinov}
\email[E-mail:]{denis@oist.jp}
\affiliation{Quantum Dynamics Unit, Okinawa Institute of Science and Technology (OIST) Graduate University, Onna, 904-0495 Okinawa, Japan}

\date{\today}

\begin{abstract}
Radio-frequency (rf) reflectometry provides a useful method for sensitive and fast detection of dynamic processes in quantum systems. We use this method to detect excitation of the quantized (Rydberg) states of the vertical motion in a system of surface-bound electrons on liquid helium. The Rydberg transition is detected as a change in the rf reflection from a resonant circuit coupled to microwave-excited electrons. It is found that the observed reflection response drastically differs from an expected state-dependent dispersive shift of the resonant frequency. To elucidate the origin of the observed response, the result is compared with an independent reflection measurement on the same electron system modulated by an electrostatic potential and with a numerical simulation using the Green's function method. The findings suggest that the observed response to the Rydberg resonance must be attributed to the in-plane motion of the photo-excited many-electron system, rather than to the Rydberg transitions of independent electrons. Our work demonstrates that rf reflectometry could be a useful tool to study dynamic processes, such as quantum ac transport and collective excitations, in this system.    

\end{abstract}     

\maketitle

\section{Introduction}
\label{sec:intro}

Condensed noble-gas elements with positive (repulsive) electron affinity, such as helium and neon, are uniquely capable of trapping electrons on their free surface~\cite{Cole1969,Shik1970,GuoMeJin}. This property provides a pristine and disorder-free environment for isolated electrons, which makes this system a highly promising platform for addressing the challenges associated with coherence of single-electron qubits~\cite{platzman1999qubits,Dykman2003,Lyon2006,Schuster2010}. Recent works have demonstrated integration of electrostatic electron traps on liquid helium and solid neon with microwave detection using a circuit quantum electrodynamics (cQED) architecture, thus allowing readout of the quantized motional states of the electronic in-plane motion and demonstration of single-qubit operations~\cite{koolstra2019_NC,zhou2022_Nature,zhou2023_NatPhys}. The quantized anharmonic states of electronic out-of-plane motion, which arise from the interaction with an image charge inside the substrate, could be also a valuable resource for qubit implementation. Such states, which are traditionally called the Rydberg states, can mediate interaction between electron spins and could be used for a non-destructive spin-state readout, as was recently suggested~\cite{kawakamiPRA2023qubits,kawakamiAPL2024review}. However, a sensitive and fast detection of the Rydberg excitation, with the typical transition frequency above 100~GHz, remains a challenge. Although a cQED architecture could be theoretically employed for such high frequencies, it presents several technical challenges, such as higher radiative losses, parasitic effects, and increased complexity of the millimeter-wave (mm-wave) transmission and coupling.~\cite{anferov2020pra,anferov2024super}. These factors currently limit the practical implementation of an efficient cQED architecture for the Rydberg state readout in this system. 

If the state of a system coupled to an electronic device can be mapped to the device impedance, radio-frequency techniques can be used for sensitive and fast state readout. In rf reflectometry, both resistive and reactive changes in the load impedance presented by a device can be measured with a high speed and accuracy by employing an ordinary 50~$\Omega$ transmission line and lumped-element impedance-matching network~\cite{schoel1998Sci,ares2016prapp,aresAPR2023}. This technique, which was primarily designed for measuring charge occupation of quantum dots (QDs) in semiconductors~\cite{reil2007apl,cass2007apl}, has flourished to become a valuable toolbox for characterization of various quantum devices and phenomena, including rf readout of semiconductor spin-qubits~\cite{vand2019natnano,urd2019natnano}, superconducting circuits~\cite{dels2006prl,john2006prb,bell2012prb}, nanomechanical resonators~\cite{scwab2007nano,ares2016rpl}, and fast thermometers~\cite{cle2003apl,pek2018prapp}. Extending these methods to other systems, such as electrons trapped on cryogenic noble-gas substrates, presents an attractive idea due to very high sensitivity of rf measurements. In particular, reminiscent of the dispersive readout in cQED, in gate-based sensing the self-resonance of an rf tank circuit is modified by the state of a charged system capacitively coupled to gate electrodes comprising the circuit. This method demonstrates an unprecedented charge sensitivity at the $\mu e/\sqrt{\textrm{Hz}}$ level~\cite{gonz2015natcomm,gonz2018prapp}. It was suggested that gate-based dispersive sensing can be used for the detection of the Rydberg transition of trapped electrons on liquid helium, as an extension of the image-charge detection technique developed earlier~\cite{kawakami2019image,kawakamiPRA2023qubits}. 


Fabrication of nano-scale traps and trapping of a single-electron on cryogenic noble-gas substrates remains a rather challenging problem, despite some notable recent progress~\cite{cast2024}. However, demonstration of rf detection in a large ensemble of microwave-excited electrons coupled to a macroscopic gate electrode should be feasible. In addition to prospects towards quantum state readout, gate-based sensing could be very useful for studying many-electron dynamics in this system, such as ac electron transport, thus complementing the conventional Sommer-Tanner (ST) measurements of the dc transport~\cite{sommerPRL1971tanner,grimesPRL1979wigner,glattliPRL1985emp,chepelianskiiNatComm2015incompressible,reesPRL2016slip}. Such a detection of the Rydberg resonance in a macroscopic ensemble of surface electrons trapped between two plates of a parallel-plate capacitor integrated in a lumped-element 120~MHz tank circuit has been recently demonstrated~\cite{kawakamiPRL2025}. However, the observed response differs drastically from what is expected from the mechanism of gate-based dispersive sensing considered earlier~\cite{kawakami2019image,kawakamiPRA2023qubits}. This calls for a more detailed investigation aimed to distinguish between different mechanisms that can lead to the observed response. Here, we report on such an investigation in an ensemble of microwave-excited electrons coupled to a lumped-element 100~MHz tank circuit. In this work, the Rydberg resonance of electrons is observed by measuring a sideband signal appearing in the rf reflection spectrum in response to the pulse modulated (PM) mm-wave excitation of electrons. Using this method, both resistive and capacitive changes in the electrical impedance of the many-electron system in response to the excitation can be observed. It is found that the measured reflection spectrum differs drastically from that expected for the dispersive frequency shift, which follows from our theoretical model. To help understanding the origin of the observed response and to distinguish between different mechanisms, we performed and analyzed an independent impedance measurement on the electrons without applied microwave excitation and subject to a modulated electrostatic potential that induced their in-plane motion. A remarkable similarity between the reflection spectra obtained by two methods strongly suggest that the observed rf response from the microwave-excited electrons originates from their collective in-plane motion induced by the resonant excitation, rather than the Rydberg transition of independent electrons which induce only their out-of-plane motion. The dependence of the reflection response on the modulation frequency and comparison with a well-understood image-charge detection of the Rydberg resonance further support this conclusion.

The paper is organized as follows. In Sec.~\ref{sec:theory}, for the sake of comparison with the experimental results, we derive the reflection spectrum for a single-mode resonator coupled to the Rydberg transition of a single electron. In Sec.~\ref{sec:setup} we describe our experimental setup and detection methods. In Sec.~\ref{sec:demod} we present our results on demodulated reflection response and make comparison with the numerical simulations. Secs.~\ref{sec:ex}~B-D present results obtained by other methods, which further support our conclusions. In Sec.~\ref{sec:disc} we discuss the main result of this work and summarize the prospects for using this method for state detection in electrons on liquid helium, as well as study of many-electron dynamics in this system.     

\section{Expected reflection response}
\label{sec:theory}

In this section, we derive an expression for the reflection response of a single-port rf resonator with the resonant frequency $\omr$ containing a single electron on liquid helium irradiated by a classical mm-wave excitation field. We assume that the frequency of the mm-wave excitation $\ommm >>\omr$ is close to the frequency $\omega_{21}$ of the transition of an electron from the ground state to the first excited Rydberg state, with the detuning $\epsilon = \omega_{21} - \ommm$. The Rabi frequency of mm-wave excitation is given by $\omega_1 = e E_\textrm{mm} z_{21}/\hbar$, where $E_\textrm{mm}$ is the electric field of mm-waves and $z_{21}$ is the transition dipole moment of an electron. The coupling of electron to a single-mode quantized resonator field $a$, with the vertically-polarized electric field $E_0$ at the location of the electron, is described by $e E_0 z(a+a^\dagger)$, where $z$ is the electron coordinate perpendicular to the surface. In the rotating frame with respect to the mm-wave excitation field, the Hamiltonian of the system can be represented as

\begin{eqnarray}
&& H/\hbar=\frac{\epsilon}{2}\sigma_z + \frac{\omega_1}{2}\sigma_x + \omr a^\dagger a + g(a + a^\dagger)\sigma_z \nonumber \\
&& - i\sqrt{\ke} (\ai a^\dagger e^{-i\omc t} - \ai^* a e^{i\omc t}),
\label{eq:H}
\end{eqnarray}

\noindent where $g=eE_0\Delta z/\hbar$ is the coupling constant, $\Delta z$ is the difference in the mean values of $z$ for the electron in the ground state and the first-excited Rydberg state, $\ke$ is the loss rate of the resonator field through the input port, and $|\ai|^2$ represents the number of photons of the driving field at the frequency $\omc$ entering the resonator per second. In the above expression, we consider only the two lowest Rydberg states described by the Pauli operators $\sigma_z$ and $\sigma_x$, and neglect all terms oscillating with frequency $\ommm$. To proceed further, it is convenient to consider the Hamiltonian in the rotating frame with respect to the resonator driving field. By applying a unitary transformation $U=\exp(-i\omc a^\dagger at)$, we obtain

\begin{eqnarray}
&& H_\textrm{rot}/\hbar=\frac{\epsilon}{2}\sigma_z + \frac{\omega_1}{2}\sigma_x + \Delta_0 a^\dagger a - i\sqrt{\ke} (\ai a^\dagger - \ai^* a )  \nonumber \\
&& + g(a e^{-i\omc t} + a^\dagger e^{i\omc t})\sigma_z.
\label{eq:Hrot}
\end{eqnarray}

\noindent where $\Delta_0 = \omr - \omc$. In the rotating frame, we are interested in the steady-state mean value of $a$, which would allow us to find the rf field reflection from the resonator using the standard input-output theory and compare it with the experimental result. Therefore, it is convenient to separate the fast motion of the coupled system during one period of oscillations $T=\omc/(2\pi)$ from the slow dynamics in the rotating frame using the Floquet theorem. For typical carrier frequencies of the rf drive $\omc >> \omega_1, g$, the slow dynamics can be described by a Floquet-gauge invariant effective Hamiltonian $H_\textrm{eff}$ found by the high-frequency expansion (HFE) in inverse powers of $\omc$~\cite{bukovAP2015}. To the second order in $\omc^{-n}$, the effective Hamiltonian is given by           

\begin{eqnarray}
&& H_\textrm{eff}/\hbar=\frac{\epsilon}{2}\sigma_z + \frac{\tilde{\omega}_1}{2}\sigma_x + \left( \Delta_0 -\frac{2g^2\omega_1}{\omc^2} \sigma_x \right) a^\dagger a  \nonumber \\
&& - i\sqrt{\ke} (\ai a^\dagger - \ai^* a ),
\label{eq:Heff}
\end{eqnarray}

\noindent where $\tilde{\omega}_1 = \omega_1 (1-2g^2/\omc^2)$. The above Hamiltonian shows that the effect of coupling of the cavity field to an externally driven two-level system is manifested as a dispersive shift of the cavity resonance proportional to the mean value of $\sigma_x$. In the presence of dissipation and dephasing, the dynamics of the system is described by a master equation 

\begin{eqnarray}
&& \frac{d\rho_\textrm{s}}{dt} = i[\rho_\textrm{s},H_\textrm{eff}] - \kappa \left( \frac{\{a^\dagger a,\rho\}}{2}- a\rho a^\dagger \right) \nonumber \\
&& - 2\gamma_\phi (\rho_\textrm{s} - \sigma_z\rho_\textrm{s}\sigma_z) - \gamma_{21} \left( \frac{\{\sigma_+\sigma_-,\rho_\textrm{s}\}}{2} - \sigma_-\rho_\textrm{s}\sigma_+  \right) \nonumber \\
&& - \gamma_{12} \left( \frac{\{\sigma_-\sigma_+,\rho_\textrm{s}\}}{2} - \sigma_+\rho_\textrm{s}\sigma_-\right), 
\label{eq:master}
\end{eqnarray}        

\noindent where $\kappa$ is the decay rate of the resonator field, $\gamma_\phi$ is the pure dephasing rate, and $\gamma_{nn'}$ is the relaxation rate of electron from $n$-th to $n'$-th Rydberg state. For an electron on liquid helium with non-quantized in-plane motion, the rates $\gamma_\phi$ and $\gamma_{nn'}$ are affected by the quasi-elastic one-ripplon scattering processes and can be obtained using the standard master equation approach by averaging over the states of the in-plane motion in thermal equilibrium at temperature $\Te$. These rates are derived in Appendix~\ref{app:relax}. Note that $\gamma_{12}=\gamma_{21} e^{-\hbar\omega_{21}/(k_B \Te)}$, in accordance with the principle of the detailed balance. Thus, for typical temperatures of the electron system $T_\textrm{e}\lesssim 1$~K, the last term in Eq.~\eqref{eq:master} can be safely neglected. Similarly, the numerical estimation shows that for typical values of the pressing electric field used in the experiment the pure dephasing rate $\gamma_\phi << \gamma_{21}/2$, thus it can be also neglected. We note that an alternative approach is to consider the coupling of the resonator field to the dressed states of the microwave-excited two level system~\cite{kawakamiPRL2025}. However, in this case one needs to derive the relaxation rates and the Bloch equations for the dressed states, which is not a trivial problem. 

Using Eqs.~\eqref{eq:Heff} and \eqref{eq:master}, the equations of motion and the steady-state solutions for the mean values can be obtained. In general, the master equation \eqref{eq:master} leads to an infinite set of coupled equations for the first-order and higher moments. To obtain an analytical expression for the reflection response, we ignore quantum correlations and factorize the second-order and higher moments. This gives a closed set of non-linear equations for the mean values of $\meana$ and $\langle \sigma_i \rangle$, $i=x,y,z$

\begin{eqnarray}
&& \frac{d\meana}{dt} = -i\Delta_0\meana + i\delta\omr \meana \meansx  - \sqrt{\ke}\ai - \frac{\kappa}{2}\meana, \nonumber \\
&& \frac{d\meansz}{dt} = \left( \tilde{\omega}_1 - 2\delta\omr \meann \right) \meansy - \gamma_{21} (1+\meansz), \nonumber \\
&&  \frac{d\meansx}{dt} = -\epsilon\meansy - \frac{\gamma_{21}}{2} \meansx, \nonumber \\
&&  \frac{d\meansy}{dt} = \epsilon\meansx - \tilde{\omega}_1\meansz +2\delta\omr\meann\meansz - \frac{\gamma_{21}}{2} \meansy, 
\label{eq:motion}
\end{eqnarray}         

\noindent where $\delta\omr = 2g^2\omega_1/\omc^2$ and $\meann = \langle a^\dagger a \rangle$ is the mean number of photons in the resonator. From the last three lines, we can find an expression for the mean values $\meansx$, $\meansy$, and $\meansz$, with the latter given by

\begin{equation}
\meansz = -\frac{1}{\left[ 1 + \frac{(\tilde{\omega}_1 - 2\delta\omr \meann)^2}{2(\epsilon^2 + (\gamma_{21}/2)^2)} \right]}.
\label{eq:sigmz}
\end{equation} 

\noindent Then, we can find the relation between $\meana$ and $\ai$ from the first equation in \eqref{eq:motion}. Finally, the reflection coefficient $\Gamma =\ao/\ai$ can be found according to the standard input-output theory from the boundary condition for a one-sided cavity~\cite{gardinerPRA1985}

\begin{equation}
\ao - \ai = \sqrt{\ke} \meana.
\end{equation}

The steady-state solution for $\meana$ and the reflection coefficient depends on the value of $\meann$. For a sufficiently small number of photons, such that $\meann << \omc^2/(4g^2)$, we obtain

\begin{equation}
\Gamma = -1 + \frac{\ke}{\frac{\kappa}{2}+i\Delta_0 - \frac{i\epsilon\tilde{\omega}_1\delta\omr}{\epsilon^2 + (\gamma_{21}/2)^2 + \tilde{\omega}_1^2/2}}.
\label{eq:Gamma}
\end{equation}

\begin{figure}[htp]
\centering
\includegraphics[width=1\linewidth]{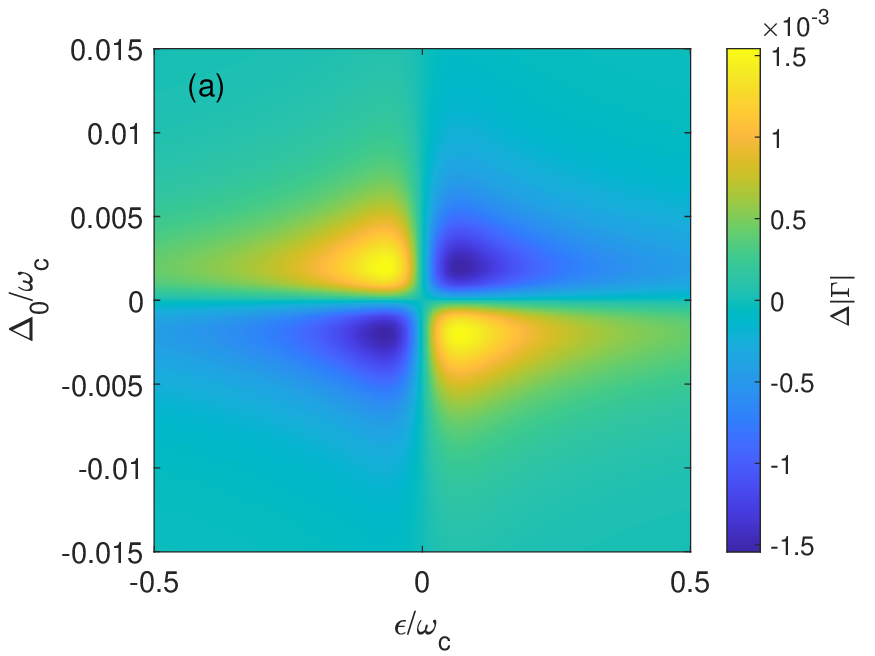}
\includegraphics[width=1\linewidth]{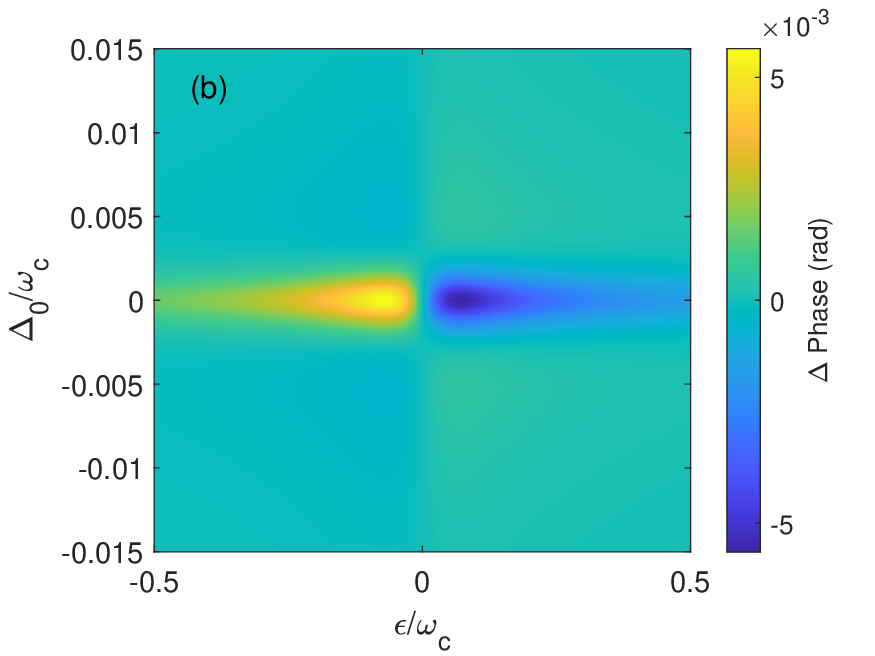}
\caption{\label{fig:1} (color online) (color online) The difference in the absolute value (a) and the phase (b) of the reflection coefficient $\Gamma$ for a resonator coupled to an electron and an empty resonator calculated using Eq.~\eqref{eq:Gamma} as a function of $\epsilon$ and $\Delta_0$ for $g=0.01\omc$, $\omega_1=0.1\omc$, $\gamma_{21} = \kappa = 0.01\omc$, and $\ke = \kappa/3$.}
\end{figure}

\noindent As an illustration, the difference between the absolute values and phase of $\Gamma$ for an electron-coupled resonator and an empty resonator calculated using Eq.~\eqref{eq:Gamma} as a function of $\epsilon$ and $\Delta_0$ for $g=0.01\omc$, $\omega_1=0.1\omc$, $\gamma_{21}=\kappa = 0.01\omc$, and $\ke = \kappa/3$ is shown in Fig.~\ref{fig:1}. As expected, the reflection shows a capacitive response due to the dispersive shift of the resonator frequency. A characteristic feature is vanishing response at zero detuning $\epsilon$. This corresponds to the vanishing mean value $\meansx$, as follows from the usual Bloch equations for a two-level system under a resonant excitation. This is a feature that should be looked for in an experiment.

For a large number of photons in the resonator $\meann >> \omc^2/(4g^2)$, the reflection response decreases and restores to that of an empty cavity at $\meann\rightarrow \infty$. This corresponds to the saturation of the two-level system $\meansz\rightarrow 0$ (see Eq.~\ref{eq:sigmz}) and vanishing of the dispersive shift, a phenomenon which is well known in cavity QED~\cite{walls2007}.  
             
\section{Setup and methods}
\label{sec:setup}

\begin{figure*}[htp]
\includegraphics[width=\textwidth,keepaspectratio]{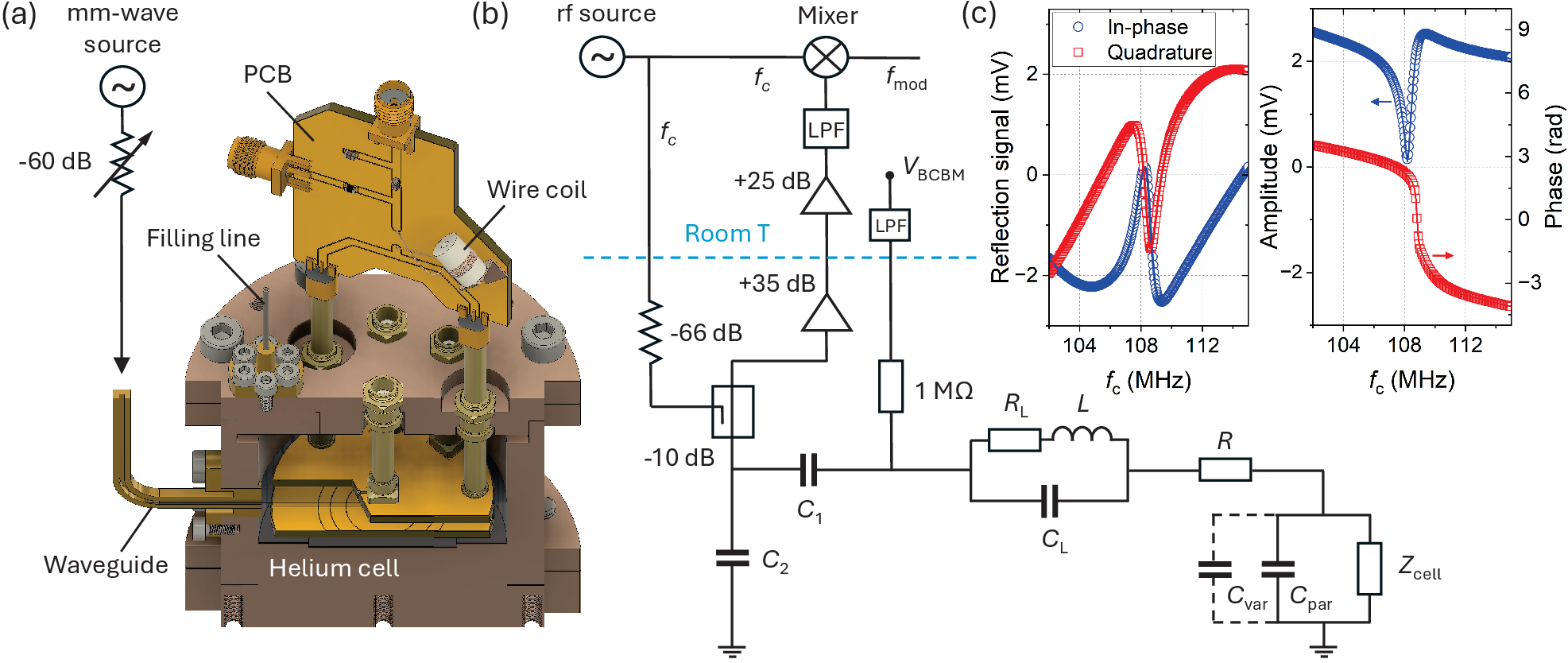}  
\caption{\label{fig:2} (color online) Experimental setup. (a) 3D rendering of the experimental cell and PCB (electrical shielding is not shown) comprising an rf device for gate-based sensing of electrons on liquid helium. (b) Circuit model of the device and measurement setup. The electrical impedance of the cell $Z_\textrm{cell}$ is represented by parallel combination of capacitance $C_\textrm{p}$ and resistance $R_\textrm{p}$. $R_\textrm{L}$ and $C_\textrm{L}$ model parasitic contributions to the impedance of the wire coil having inductance $L$, while $C_\textrm{par}$ models the parasitic capacitance of PCB and electrical connections in the cell. The effective resistance $R$ represents other losses in the circuit. The amplitude-modulated reflection from the device is amplified by a cryogenic amplifier followed by a room-temperature amplifier (Fairview Microwaves FMAM3311) and low-pass filter (Mini-Circuits LSP-250+) and demodulated by a mixer (Mini-Circuits ZEM-2B+). Alternatively, the signal at the output of the cryogenic amplifier can be measured by a signal or vector analyzer. (c) In-phase and quadrature components (left panel) and amplitude and phase (right panel) of the reflection signal measured by an rf lock-in amplifier at $T=100$~mK. Solid lines show fitting as described in the text.}
\end{figure*}

The experimental setup used for trapping of electrons on the surface of liquid helium is similar to that used in Ref.~\cite{kawakamiPRL2025}. A vacuum-tight cylindrical cell attached to the mixing chamber of a dilution refrigerator contains two circular conducting plates separated by a distance $D=2$~mm, thus forming a parallel-plate capacitor (see Fig.~\ref{fig:2}(a)). Each plate, having a diameter of 35.5~mm, consists of four concentric electrodes separated by three circular gaps (width 0.2~mm) with diameters 11.9, 16.9, and 20.9~mm. The most outer electrode is permanently grounded for each plate, while independent dc bias voltages can be applied to the three other concentric electrodes, which we refer to as the center (BC), middle (BM) and guard (BG) electrodes on the bottom plate, and the center (TC), middle (TM) and guard (TG) electrodes on the top plate. The electrical connection to each electrode is provided through the hermetic SMP connectors at the top of the cell via six SMP bullet adapters, as shown in Fig.~\ref{fig:2}(a). The circuit model of the device and measurement setup is shown in Fig.~\ref{fig:2}(b). For gate-base sensing of electrons, a wire coil is connected in series with the center and middle electrodes of the bottom plate, thus forming a lumped-element tank circuit whose resonance frequency $\omr$ is determined by the coil inductance $L$, parasitic capacitance of the coil $C_\textrm{L}$, parasitic capacitance of PCB connections $C_\textrm{par}$, and the impedance of the cell $Z_\textrm{cell}$ which can be modeled as a parallel combination of capacitance $C_\textrm{p}$ and resistance $R_\textrm{p}$. The coil is made of 0.1~mm-diameter copper wire wound on a 3.6-mm-diameter PTFE cylinder with 9 turns and showed the inductance $L=777$~nH and self-resonance at $385$~MHz at room temperature. It is mounted on a shielded PCB and connected to the electrodes inside the cell via two SMP connectors (each one for center and middle electrodes, respectively), as shown in Fig.~\ref{fig:2}(a). The PCB also contains a 1~M$\Omega$ resistor for the dc biasing (with voltage $V_\textrm{BCBM}$) of the central and middle electrodes of the bottom plate, and a capacitance divider comprised of $C_1=10$~pF and $C_2=56$~pF that matches the device impedance to 50~$\Omega$ impedance of the transmission line~\cite{ares2016prapp}. The circuit losses are represented by the coil resistance $R_\textrm{L}$ and an effective series resistance $R$. In a later experiment, a voltage-tunable varactor $C_\textrm{var}$ (Macom MA46H204-1056) was added in parallel with the impedance of the cell, as indicated by dashed lines in Fig.~\ref{fig:2}(b), to calibrate the capacitance sensitivity of the setup, as described in Appendix~\ref{app:sideband}.

For rf reflectometry measurements, a carrier signal at the frequency $f_\textrm{c}=\omc/(2\pi)$ (with a typical power of -3~dBm used in this experiment) from a room-temperature rf source is fed into the circuit through an attenuated (-66 dB in total) cryogenic 50~$\Omega$ coaxial line via a -10~dB directional coupler (Mini-Circuits ZEDC-15-2B) attached to the mixing chamber plate. The reflected signal is directed by the coupler to a cryogenic low-noise voltage amplifier (Cosmic Microwave Technology, CITLF1) and measured at the room temperature. Fig.~\ref{fig:2}(c) (left panel) shows the in-phase (open circles) and quadrature (open squares) components of the reflection signal recorded using an rf lock-in amplifier (Stanford Research SR844) with the cell connected to the PCB and cooled down to $T=100$~mK. The amplitude and phase of the measured signal are shown on the right panel. Following a standard method, the observed asymmetric lineshape is fitted (solid lines) by taking into account a phase and amplitude distortion due to impedance mismatching~\cite{prob2015rsi}. From this fitting procedure, the resonant frequency $\omr/(2\pi)=108.4$~MHz and the internal and external quality factors $Q_\textrm{i}\approx 228$ and $Q_\textrm{e}\approx 191$, respectively, of the device are obtained, thus showing that the tank circuit is slightly overcoupled to the feedline.

\begin{figure}[htp]
\includegraphics[width=\columnwidth,keepaspectratio]{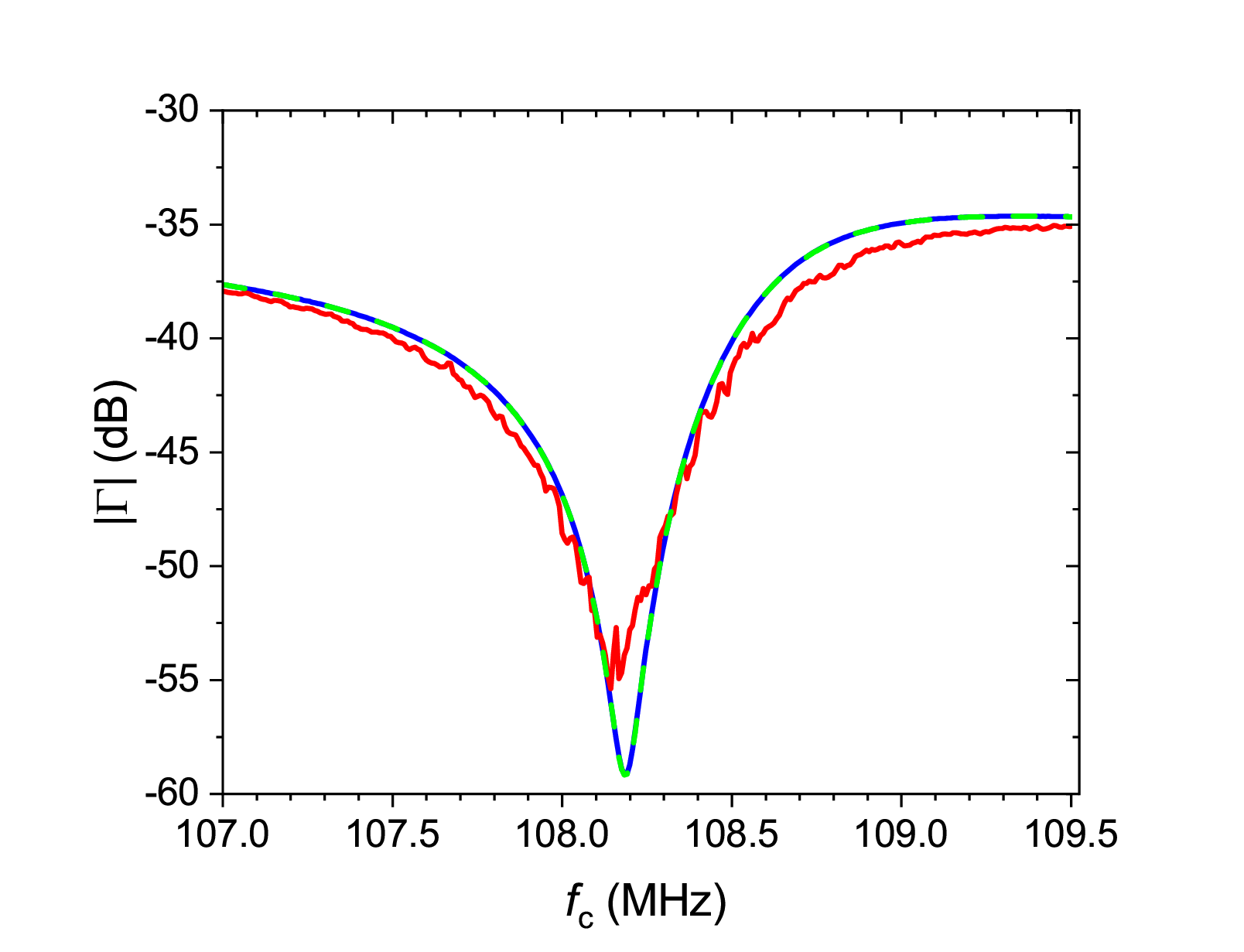}  
\caption{\label{fig:3} (color online) Exemplary reflection spectra (solid lines) taken using VNA before (blue) and after (red) charging the surface of liquid helium with electrons. The dashed line is the reflection spectrum taken with electrons confined by negative guard potentials.}
\end{figure}

After the cell is cooled down below 1~K, the liquid helium is condensed into the cell and the liquid level is set approximately midway between the bottom and top plates of the parallel-plate capacitor, as determined by measuring the capacitance between the guard electrodes of the bottom and top plates using a capacitance bridge (Andeen-Hagerling 2700A). Electrons are produced by the thermionic emission from a tungsten filament mounted on the top plate close to the guard electrode, while a positive dc bias voltage of $V_\textrm{BCBM}=20$~V is applied to the central and middle electrodes of the bottom plate ($V_\textrm{BCBM}=V_\textrm{BC}=V_\textrm{BM}$), while all other electrodes are grounded. The areal density of surface electrons after charging corresponds to the complete screening of the electric field above the charged surface of liquid. It is observed that the reflection spectrum of the device becomes very noisy, presumably due to variation in the impedance of the cell caused by fluctuations of the charged surface of liquid. An example of reflection spectra taken before and after the electron deposition are shown in Fig.~\ref{fig:3}. In order to stabilize the electron system, the dc voltage $V_\textrm{BCBM}$ is increased to 30~V, while the guard electrodes of the top and bottom plates are set to $V_\textrm{BGTG}=-60$~V ($V_\textrm{BGTG}=V_\textrm{BG}=V_\textrm{TG}$). It is observed that under such conditions the refection spectrum nearly coincides with the reflection spectrum without electrons, see Fig.~\ref{fig:3}.

The electrons are excited by mm-wave radiation at the frequency $f_\textrm{mm}=\ommm/(2\pi)$ transmitted from a room-temperature source with output power of about 10~mW through a calibrated variable attenuator (maximum attenuation below -60~dB) and a transmission waveguide coupled to the cell (see Fig.~\ref{fig:2}(a)). The transition frequency of electrons $\omega_{21}$ is adjusted to match the range of the used values of $\ommm$ by applying a positive voltage $V_\textrm{TCTM}$ to the central and middle electrodes of the top plate ($V_\textrm{TCTM}=V_\textrm{TC}=V_\textrm{TM}$) which shifts the energy levels of the Rydberg states by the Stark effect~\cite{CollinPRL2002}. As described in Appendix~\ref{app:WS}, it is found that upon applying positive $V_\textrm{TCTM}$ the surface of liquid is partially discharged such that the areal density of electrons $n_\textrm{s}$ corresponds to the complete screening of the electric field $E_\perp \approx (V_\textrm{BCBM}-V_\textrm{TCTM})/D$ due to the applied bias voltages. The corresponding values of $\ns$ are confirmed by measuring the temperature of the transition of electrons to the Wigner solid phase (see Appendix~\ref{app:WS}).

To detect small changes in the rf reflection due to excitation of electrons, the mm-wave radiation is pulse-modulated at the frequency $\omm$ and the reflected signal is demodulated at room temperature to an ac signal at the frequency $\omm$ by mixing it with a local oscillator at the carrier frequency $\omc$ (see Fig.~2(b)). The demodulated signal is measured by an ordinary lock-in amplifier referenced at the modulation frequency $\omm$.  Assuming that the capacitive and resistive changes in the cell impedance $Z_\textrm{cell}$ are given by $\delta C_\textrm{p}\sin (\omm t+\phi_0)$ and $\delta R_\textrm{p}\sin (\omm t+\phi_0)$, respectively, the amplitude of the lock-in measured signal $V_\textrm{s}$ depends on $\delta C_\textrm{p}$ and $\delta R_\textrm{p}$ according to

\begin{equation}
V_\textrm{s} \propto \textrm{Re}\left[\left(\frac{d\Gamma}{dC_\textrm{p}}\right)\delta C_\textrm{p} + \left(\frac{d\Gamma}{dR_\textrm{p}}\right) \delta R_\textrm{p} \right].
\label{eq:lockin}
\end{equation}

\noindent It is expected that the capacitive change produces a dispersive voltage response that changes sign with varying carrier frequency and vanishes at the resonant frequency of the rf circuit, while the resistive change produces an absorptive voltage response which is maximum at the resonance frequency~\cite{aresAPR2023}. This is confirmed by our simulation of the reflection response using the circuit model of our setup, as described in Appendix~\ref{app:reflection}. As an alternative detection scheme, the rf response of the system to the modulated Rydberg excitation can be measured using a spectrum analyzer (SA) by observing the sidebands appearing in the reflection power spectrum at frequencies $\omc\pm \omm$. In this case, the sideband amplitude is proportional to $|(d\Gamma/C_\textrm{p})\delta C_\textrm{p} + (d\Gamma/dR_\textrm{p})\delta R_\textrm{p}|^2$. This method also provides a convenient way to quantify the sensitivity of the Rydberg-resonance detection in terms of the signal-to-noise ratio (SNR) for a given bandwidth by measuring (in dB) the height of the sideband from the noise floor~\cite{schoel1998Sci}, as described in Appendix~\ref{app:sideband}.

In order to estimate the detected response in terms of the excited state population, the Rydberg transition is independently measured by the image-charge detection method using a resonant image-current amplifier developed earlier~\cite{mishaJLTP2023amp}. This cryogenic amplifier consists of a superconducting helical resonator (not shown in Fig.~\ref{fig:2}(a)), which is connected to the central electrode of the top plate, followed by a high-input-impedance two-stage voltage amplifier. As described in details previously~\cite{mishaJLTP2023amp}, a large real impedance of the resonator transforms the image-current signal induced in the central electrode by the excited electrons into a voltage signal. This signal is amplified and detected at room temperature by another lock-in amplifier referenced at the modulation frequency of the mm-wave excitation, which must coincide with the resonance frequency of the resonator ($f_\textrm{res}=1.20483$~MHz). From the magnitude of the measured current signal, the excited-state population can be determined, as described in Sec.~\ref{sec:image}.       

\section{Experimental Results}
\label{sec:ex}
\subsection{Response to the Rydberg excitation}
\label{sec:demod}

\begin{figure}[htp]
\includegraphics[width=\columnwidth,keepaspectratio]{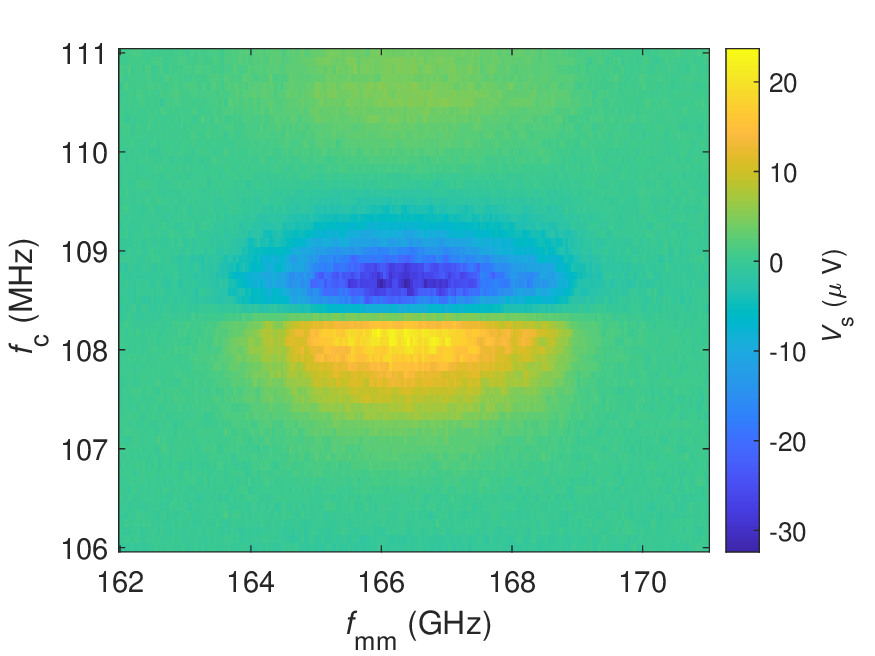}  
\caption{\label{fig:4} (color online) A color map of the measured demodulated reflection signal versus the mm-wave frequency $f_\textrm{mm}$ and the rf carrier frequency $f_\textrm{c}$ obtained for PM mm-wave excitation at the modulation frequency of 1~kHz and confining voltages $V_\textrm{BCBM}=30$~V, $V_\textrm{TCTM}=18$~V and $V_\textrm{BGTG}=-60$~V corresponding to the electron density profile given in Fig.~\ref{fig:11} in Appendix~\ref{app:green}. } 
\end{figure}

First, we present our measurement results for the demodulated response of the rf reflection signal due to PM mm-wave excitation of electrons. In the experiment, the mm-wave frequency $\ommm$ is varied to match the transition frequency of electrons $\omega_{21}$ corresponding to their excitation from the ground state to the first excited Rydberg state, while the rf carrier frequency $\omc$ is varied to tune refection in resonance with the tank circuit. Fig.~\ref{fig:4} shows a color map of the demodulated voltage signal measured by a lock-in amplifier with the measurement bandwidth of about $0.1$~Hz (the settling time 4 seconds) versus $f_\textrm{mm}$ and $f_\textrm{c}$. The data are taken for an electron system confined above the central and middle electrodes of the bottom plate with confining voltages $V_\textrm{BCBM}=30$~V, $V_\textrm{TCTM}=18$~V, and $V_\textrm{BGTG}=-60$~V, corresponding to the calculated electron density profile shown in Fig.~\ref{fig:11} in Appendix~\ref{app:green}. The mean value of the areal density of electrons was confirmed by observation of the Wigner solid transition of electrons, as described in Appendix~\ref{app:WS}. The incident mm-wave power is adjusted by setting the variable attenuator at the mm-wave source (see Fig.~\ref{fig:2}(a)) at -40~dB, which corresponded to the highest input power used in this experiment. The signal due to the Rydberg transition centered around $166.5$~GHz is clearly observed.  This frequency is in a good agreement with the transition frequency $\omega_{21}/(2\pi)$ expected for the Stark-shifted Rydberg energy levels corresponding to the vertical electric field $E_\perp\approx (V_\textrm{BCBM}-V_\textrm{TCTM})/D=6$~kV/m~\cite{CollinPRL2002}. 

\begin{figure}[htp]
\centering
\includegraphics[width=1\linewidth]{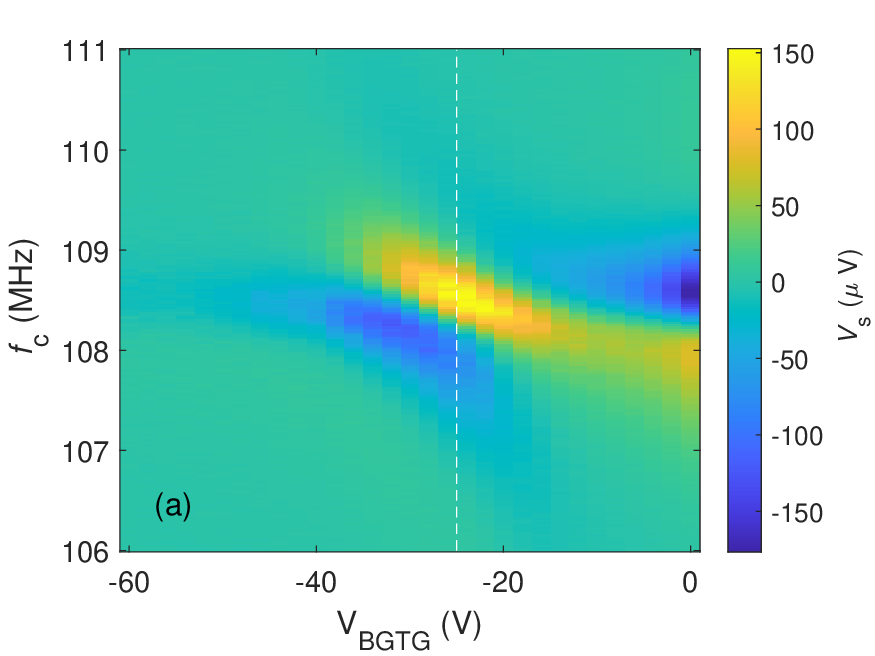}
\includegraphics[width=1\linewidth]{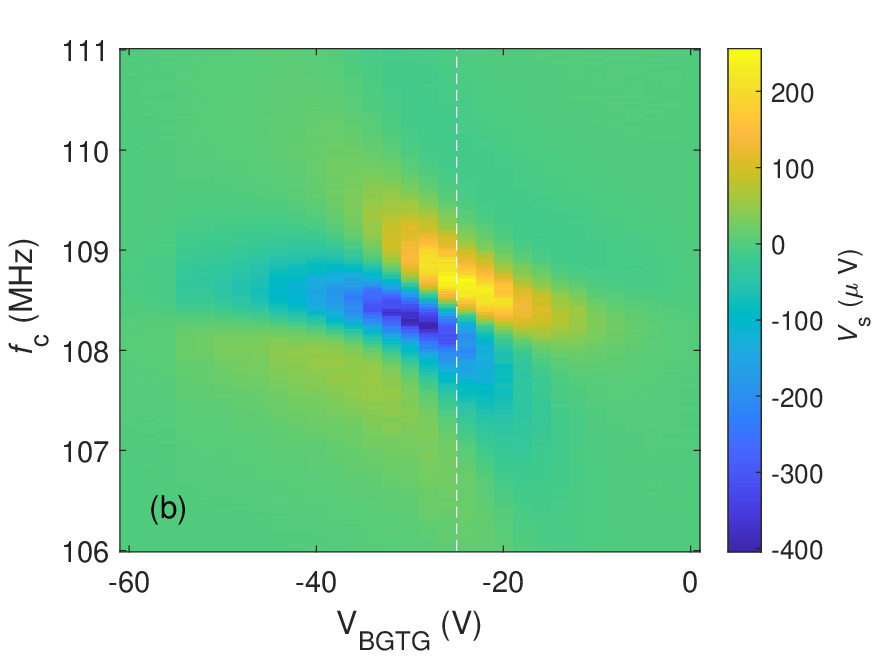}
\caption{\label{fig:5} (color online) (a) A color map of the demodulated reflection signal measured without mm-wave excitation and applying an ac modulation voltage 3~V$_{\textrm{p-p}}$ to the guard electrodes versus the dc guard voltage $V_\textrm{BGTG}$ and the rf carrier frequency $f_\textrm{c}=\omc/(2\pi)$. (b) A similar color map but obtained by removing the ac voltage modulation and applying a resonant mm-wave excitation at the frequency of 166.5~GHz. The vertical dashed line in both panels indicates the voltage at which the lowest resonant mode of the plasmon oscillations nearly coincides with the resonant frequency of the tank circuit, see Fig.~\ref{fig:12} in Appendix~\ref{app:green}.}
\end{figure}

The reflection signal in Fig.~\ref{fig:4} changes sign with varying carrier frequency $f_\textrm{c}$ and vanishes at the resonant frequency of the rf circuit. As described in Sec.~\ref{sec:setup}, this corresponds to a purely capacitive change in the cell's impedance proportional to $\textrm{Re}(d\Gamma/dC_\textrm{p})$. At the same time, the dependance of the response on the detuning from the Rydberg resonance $\epsilon = \omega_{21}-\ommm$ is drastically different from the dispersive shift of the resonator frequency described in Section~\ref{sec:theory}. Contrary to the expected reflection response shown in Fig.~\ref{fig:1}, the response in Fig.~\ref{fig:4} is largest at zero detuning. This clearly indicates that the observed response originates from a mechanism different from that considered in Sec.~\ref{sec:theory}. Note that the latter considers only the vertical motion of an electron due to the photo-excited transitions between the out-of-plane Rydberg states. However, it is expected that the mm-wave excitation also causes an in-plane motion of the many-electron system. For example, an in-plane motion of electrons in response to the local heating of electrons by the mm-wave excitation has been recently observed~\cite{KostylevPRL2021}. This effect was attributed to a thermoelectric transport of electrons but is not entirely understood. Fortunately, it is also easy to excite the in-plane motion of the many-electron system on liquid helium by a purely electrostatic method. For the sake of comparison with the effect of photo-excitation, we repeated the experiment by removing the mm-wave excitation and applying an ac voltage to the guard electrodes at the frequency equal to the modulation frequency $\omm$ of the PM excitation. It is expected that such an ac voltage will introduce modulation of the confining potential, therefore the surface area occupied by the electron system,  which should cause a predictable change in the cell's impedance. Figure~\ref{fig:5}(a) shows the demodulated reflection response obtained by applying an ac voltage with the peak-to-peak voltage of $3$~V at the modulation frequency $\omm/(2\pi)=1$~kHz versus the dc voltage $V_\textrm{BGTG}$ and the rf carrier frequency $f_\textrm{c}$. For the sake of comparison, a similar color map of the reflection response measured with an applied resonant mm-wave excitation at the frequency $f_\textrm{mm}=166.5$~GHz is shown in Fig.~\ref{fig:5}(b). A remarkable similarity between these two plots indicates that the rf reflection response to the mm-wave excitation of electrons observed in the experiment originates from the in-plane motion of the electron system, which causes variation in the impedance components $C_\textrm{p}$ and $R_\textrm{p}$ of the experimental cell.     

It is found that the reflection response strongly varies with the confining potential due to the negative guard voltage $V_\textrm{BGTG}$. At $V_\textrm{BGTG}= -60$~V, the system shows a predominantly capacitive response similar to the one shown in Fig.~\ref{fig:4} (see also Fig.~\ref{fig:14} in Appendix~\ref{app:reflection}). A drastic change in response is observed at the guard voltage around $-25$~V, which is indicated by a dashed line in Figs.~\ref{fig:5}(a,b). The reflection signal is strongly enhanced and shows a mixed capacitive/resistive response. It is found that such a response occurs due to excitation of a resonant mode of the plasmon oscillations in the many-electron system by the rf signal at the frequency close to the resonant frequency of the rf circuit. For the calculated density distribution (see Fig.~\ref{fig:11} in Appendix~\ref{app:green}) and assuming a non-degenerate 2D electron gas, the lowest-frequency resonant mode of the plasmon oscillations corresponds to the frequency $\omp/(2\pi)=77.3$~MHz. This is much lower than the result shown in Figs.~\ref{fig:5}(a,b) where the plasmon resonance occurs at the resonator frequency of about 108~MHz. However, as shown in Appendix~\ref{app:WS}, at the temperature of the experiment $T=100$~mK the electron system forms the Wigner solid (WS) and, therefore, is strongly coupled to the medium vibrations (polaronic dimples). Such a coupling can strongly affect the plasmon dispersion relation and modify the frequency of its resonant modes. In this experiment, the rf frequency $f_\textrm{c}$ is much higher than the ripplon frequency $f_{g_1}\approx 3$~MHz corresponding to the first reciprocal lattice vector of WS. Therefore, we speculate that the plasmon resonance observed in Figs.~\ref{fig:5}(a,b) might correspond to a fast optical branch of the coupled plasmon-ripplon oscillations with the frequency $\sqrt{\omf^2+\omp^2}$, where $\omf$ is the frequency of oscillations of a single electron in the potential created by a polaronic dimple ~\cite{monarkha2004}.       

\begin{figure}[htp]
\includegraphics[width=\columnwidth,keepaspectratio]{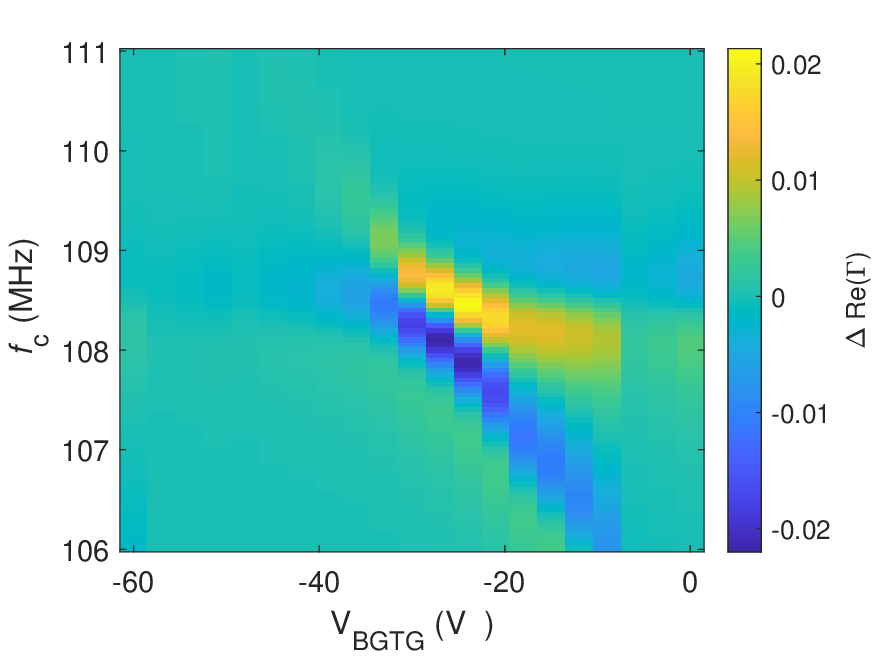}
\caption{\label{fig:6} (color online) A color map of the numerically simulated reflection response of the experimental cell containing electrons to an ac modulation voltage of 3~V$_{\textrm{p-p}}$ applied to the guard electrodes. The details of calculations are described in the text.} 
\end{figure}          

In order to quantify the variations $\delta C_\textrm{p}$ and $\delta R_\textrm{p}$ due to the applied voltage modulation, we carried out a numerical simulation to calculate the reflection response and compared it with the data presented in Fig.~\ref{fig:5}(a). We used the Green function method to calculate the complex impedance of our experimental cell containing the surface electrons and found their reflection response by analyzing the equivalent electrical circuit of the cell (see Appendices \ref{app:green} and \ref{app:reflection} for details). In this calculation, we assumed a phenomenological expression for the ac conductivity of electrons in the form

\begin{equation}
\sigma = \frac{e^2 n_\textrm{e}}{m_\textrm{e}} \frac{1}{\left[ \nu + i(\omc - \omg^2/\omc)\right]},
\label{eq:cond}
\end{equation}

\noindent where $\nu=9.5\times 10^6$~s$^{-1}$ is the momentum relaxation rate for a non-degenerate electron gas at $T=100$~mK. For $\omc >> \nu$, such an expression gives a gaped dispersion relation of plasmon oscillations. The gap frequency $\omg/(2\pi)=76$~MHz is chosen as an adjustable parameter to reproduce the resonant plasmon mode observed in Fig.~\ref{fig:5}. It is instructive to compare this value with the frequency $\omf$. The latter can be estimated from an expression $\omf^2=\sum_\textbf{g} (v_g^2/2)F_g$, where $v_q=\sqrt{n_s/(m_\textrm{e}\alpha)}V_q$, $V_q$ is the matrix element of the electron-ripplon interaction given by the last two terms in Eq.~\eqref{eq:Uqexp}, $F_g$ is the Debye-Waller factor, and sum is over all reciprocal lattice vectors $\textbf{g}$~\cite{monarkha2004}. In the vicinity of the WS melting temperature of $170$~mK, the main contribution comes from the lowest reciprocal lattice vectors $\textbf{g}_1$ (there are six such vectors), which results in $\omf/(2\pi)\lesssim 10$~MHz. At lower temperatures, the contribution from $g_n>g_1$ should be also taken into account. Surprisingly, the value of the adjustable parameter $\omg$ is significantly higher than $\omf$. However, a full account of the medium response in the electron conductivity could be a rather formidable task and requires more theoretical considerations. In the present work, we use a simple phenomenological model presented by Eq.~\eqref{eq:cond}.          

Using the above model, we calculated the complex impedance of the experimental cell with electrons as a function of the carrier frequency $\omc$ and for two different values of the confining guard potential $V_\textrm{BGTG}^{(1)}$ and $V_\textrm{BGTG}^{(2)}$, with $V_\textrm{BGTG}^{(2)}-V_\textrm{BGTG}^{(1)}=1.5$~V, that is half of the peak-to-peak voltage modulation used to obtained data shown in Fig.~\ref{fig:5}(a). The reflection coefficient $\Gamma$ is calculated using the model circuit of the experimental setup shown in Fig.~\ref{fig:2}(b) (see Appendix~\ref{app:reflection} for details). The unknown circuit parameters are adjusted to fit the experimental response at $V_\textrm{BGTG}=-60$~V, as shown by Fig.~\ref{fig:14} in Appendix~\ref{app:reflection}.  It is numerically found that the applied modulation causes the capacitance and resistance changes of $\delta C_\textrm{p} = \pm 40$~aF ($|\delta C_\textrm{p}/C_\textrm{p}|\approx 2\times 10^{-5}$) and $\delta R_\textrm{p}= \mp 0.17$~M$\Omega$ ($|\delta R_\textrm{p}/R_\textrm{p}|\approx 0.08$), respectively, with a corresponding change in the electron pool radius of $\pm 0.02$~mm. The calculations are repeated for several values of $V_\textrm{BGTG}^{(1)}$ in the range from -60 to 0~V in order to reconstruct a measured color map shown in Fig.~\ref{fig:5}(a). Fig.~\ref{fig:6} shows a color map of the difference between the real parts of $\Gamma$ calculated for $V_\textrm{BGTG}^{(2)}$ and $V_\textrm{BGTG}^{(1)}$. For different values of  $V_\textrm{BGTG}$, the radius of the electron pool was manually adjusted to keep the number of electrons fixed at $N_\textrm{e}\approx 8.65\times 10^7$ within 1\% accuracy. According to Eq.~\eqref{eq:lockin}, the calculated difference $\Delta\textrm{Re}(\Gamma)$ is proportional to the demodulated reflection response measured in the experiment. The agreement between the experimental data in Fig.~\ref{fig:5}(a) and numerical simulation in Fig.~\ref{fig:6} is quite good, thus indicating that the results obtained from our simulations are reliable. By comparing the experimental response obtained by two methods and shown in Figs.~\ref{fig:5}(a) and \ref{fig:5}(b), we can conclude that the mm-wave excitation of electrons likely leads to their in-plane motion that modifies the cell's impedance, thus causing the observed reflection response. Possible mechanisms of such in-plane motion are discussed in Sec.~\ref{sec:disc}.

\subsection{Comparison with the image charge detection}
\label{sec:image}

\begin{figure}[htp]
\includegraphics[width=\columnwidth,keepaspectratio]{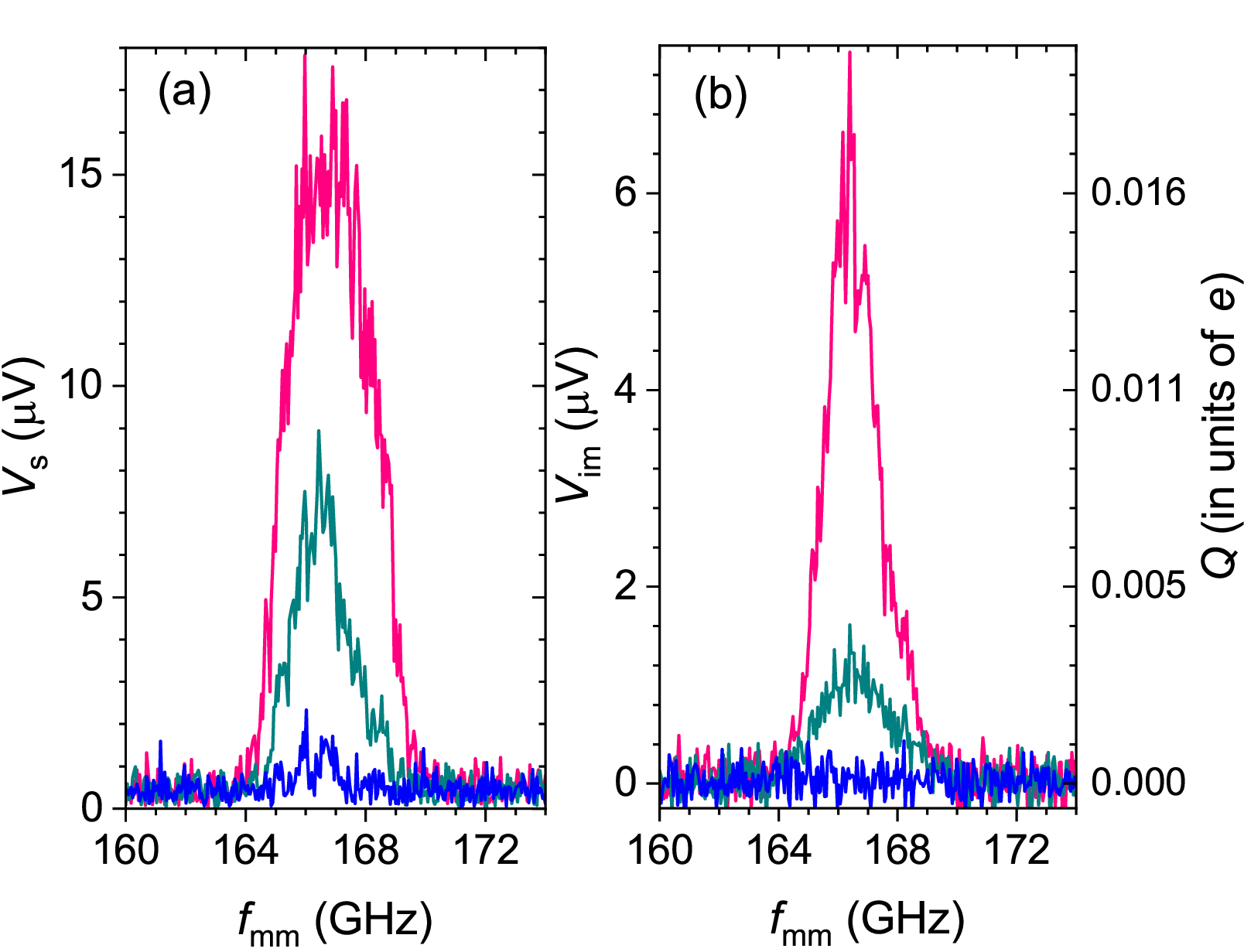}  
\caption{\label{fig:7} (color online) The demodulated rf reflection signal (a) and the image-charge detection signal (b) measured as a function of the mm-wave frequency $f_\textrm{mm}$ for three different values of the incident mm-wave power controlled by the variable attenuator. The resonance signals with progressively decreasing amplitude correspond to -40, -50 and -60~dB attenuation. The right vertical axis in panel (b) represents the image charge $Q$ (in units of the elementary charge $e$) induced at the central electrode of the top plate by the excited electrons, which is calculated from the measured voltage signal (left vertical axis), as described in the text.} 
\end{figure}

In order to learn more about the mechanism of the reflection response caused by the Rydberg excitation of the electron system, we made comparison between the rf reflection response and the image-charge response using the method described in Section~\ref{sec:setup}. The image-charge detection measures a voltage signal $V_\textrm{im}$ proportional to the image current induced by the excited electrons at the center electrode of the top plate. For this experiment, the electrons were confined between the central electrodes of the top and bottom plates by applying a strong negative potential $V_\textrm{TM}=-60$~V to the middle electrode of the top plate. Fig.~\ref{fig:7} shows comparison between the demodulated reflection signal $V_\textrm{s}$ and the image-charge signal $V_\textrm{im}$ measured as a function of the mm-wave frequency $\ommm$ for three different values of the incident power controlled by the variable attenuator (-40, -50 and -60~dB). The reflection signal (a) is measured with the carrier frequency $f_\textrm{c}=108$~MHz and PM excitation at the modulation frequency $\omm/(2\pi)=1$~kHz, while the image-charge signal (b) is measured with the PM excitation at the modulation frequency $\omm/(2\pi)=1.20182$~MHz, as described in Sec.~\ref{sec:setup}. The Rydberg resonance peak centered around 166.5~GHz is clearly observed in both cases. The peak amplitude decreases with decreasing mm-wave power, as expected. Surprisingly, the Rydberg resonance is still detected at the lowest power corresponding to -60~dB attenuation by the rf reflectometry method, while it is too small to be detected by the image-charge method. Also, the transition line shape is noticeably different for the two detection methods. The line shape obtained by the rf reflectometry method appears to be broader than the line shape of the same transition obtained using the image-charge method. 

In order to estimate the image charge induced at the detection electrode by the excitation of electrons, we recalculate the measured voltage signal $V_\textrm{im}$ in terms of the image current $I_\textrm{im}$ induced by the excited electrons using the transimpedance gain of our detection setup $I_\textrm{im}/V_\textrm{im}=3.2$~nA/V, as was previously determined~\cite{mishaJLTP2023amp}. Then, we can find the corresponding variation of the image charge due to the excitation of electrons according to $Q=I/\omm$. The corresponding values are shown on the right vertical axis of Fig.~\ref{fig:7}(b) in the units of the elementary charge $e$. This estimate can be compared with a calculated image charge for a given mm-wave excitation rate in order to make an estimate for the excited state population and the temperature of the electron system, as described in detail in Sec.~\ref{sec:disc}.

\subsection{Dependence on mm-wave power}
\label{sec:power}

\begin{figure}[htp]
\includegraphics[width=\columnwidth,keepaspectratio]{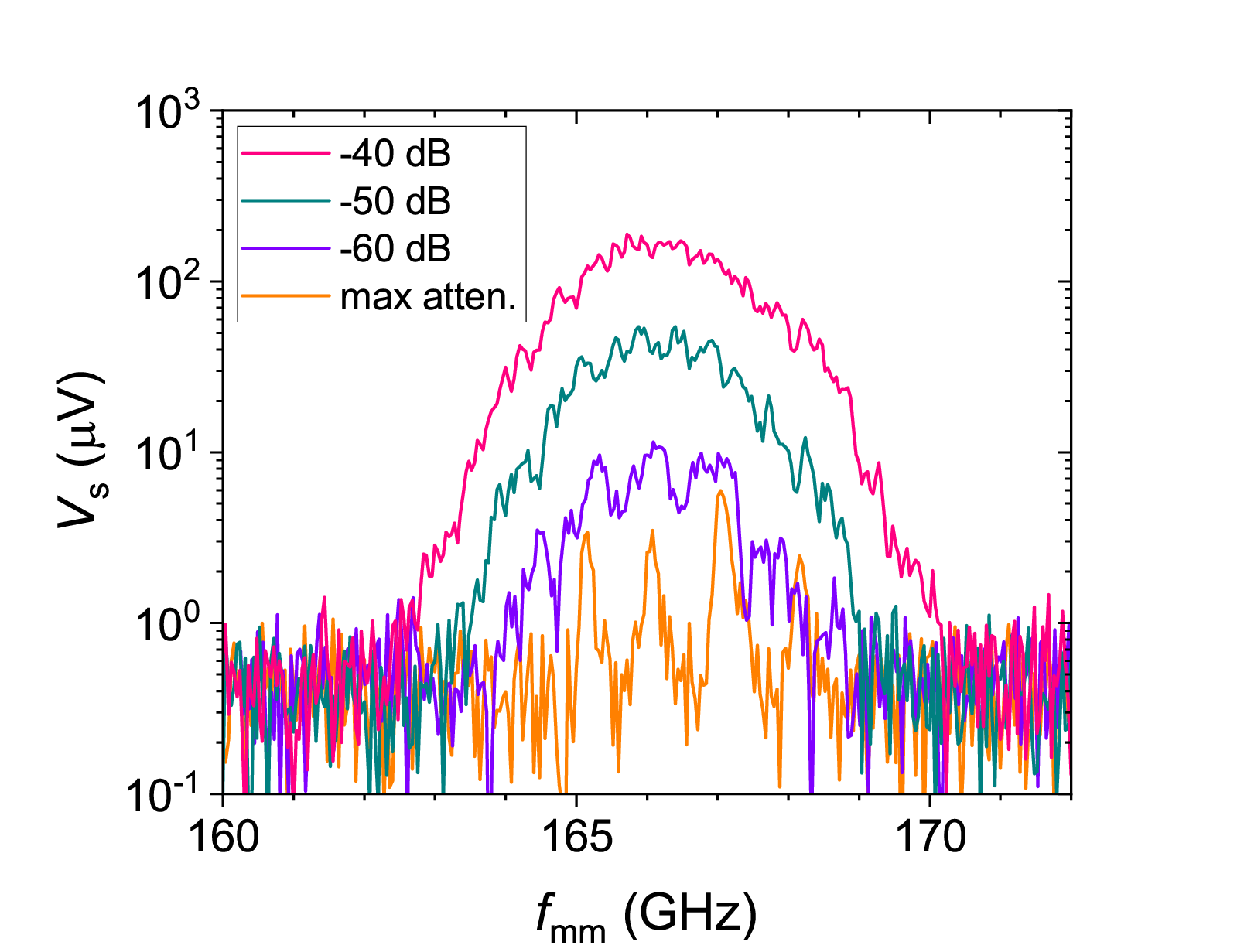}  
\caption{\label{fig:8} (color online) The log plot of the demodulated reflection signal measured for several values of the incident mm-wave power controlled by the attenuator settings, as indicated in the legend.}
\end{figure}

We also investigated the dependence of the reflection response due to photo-excitation of electrons on the mm-wave power. The measurements were done close to the plasmon resonance where the reflection signal is significantly enhanced, see Section~\ref{sec:demod}. Fig.~\ref{fig:8} shows the demodulated signal measured for several values of the incident mm-wave power. The incident power is controlled by a variable attenuator placed at the output of the room-temperature mm-wave source. Although the attenuator is factory-calibrated in the range from 0 to -60 ~dB, it was founds that the maximum attenuation, which corresponds to the fully closed position of the adjusting nob, was substantially below -60~dB. Therefore, in what follows we refer to this setting as a maximum attenuation without specifying the value. Note that the voltage signal $V_\textrm{s}$ is plotted in the log scale in order to highlight the signal observed at the maximum attenuation of the mm-wave power. Remarkably, the system response is still observed even at such a low power and represents a set of peaks equally separated in frequency by about 1~GHz. In order to elucidate origin of these peaks, we measured the demodulated response at the lowest mm-wave power by varying the pressing electric field $E_\perp$ exerted on the electrons, as described in Appendix~\ref{app:stark}. As described earlier, such field causes the Stark shift of the Rydberg energy levels of the electrons, thus allowing to investigate the Rydberg resonance in a wider radiation frequency range. Surprisingly, we observed no variation in the position of this peaks with respect to the frequency by varying the pressing field. This indicates that these peaks originate from the properties of the experimental cell, rather than the electron system.

\subsection{Dependence on modulation}
\label{sec:side}

Finally, the reflection response of the system to the modulated Rydberg excitation was measured by a method similar to the one used in Ref.~\cite{kawakamiPRL2025}, that is by observing the sidebands appearing in the power reflection spectrum at the frequencies $\omc\pm \omm$, as described in Sec.~\ref{sec:setup}. In this experiment, the wire coil was connected only to the central electrode of the bottom capacitor plate and a lower density of electrons of about $5\times 10^{10}$~cm$^{-2}$ was used by charging the surface with a dc bias voltage of $V_\textrm{BC}=1$~V applied to the central electrode. The solid line in Fig.~\ref{fig:9} shows the sideband signal-to-noise ratio (in dB) as defined in Appendix~\ref{app:sideband} measured by applying the PM mm-wave excitation at the maximum power corresponding to attenuator setting of -40~dB. In this experiment, the frequency of mm-wave excitation was fixed at 165~GHz, while the Rydberg transition of electrons was tuned in resonance via the Stark shift by varying the pressing electric field $E_\perp = V_\textrm{BC}/D$ (all other electrodes are kept at ground potential). As expected, the reflection response to the PM excitation has a similar lineshape as the one shown in Fig.~\ref{fig:8}. For the sake of a cross-check, we also measured the response by applying the frequency-modulated (FM) mm-wave excitation, which was used in Ref.~\cite{kawakamiPRL2025}. Solid squares in Fig.~\ref{fig:9} show the sideband amplitude measured by applying FM excitation at $165$~GHz with the frequency deviation $\Delta f_\textrm{mm} = 500$~MHz at the repetition rate of 1.2~kHz. Naturally, the line shape is given by the absolute value of the derivative of the line shape measured by the PM method.            

\begin{figure}[htp]
\includegraphics[width=\columnwidth,keepaspectratio]{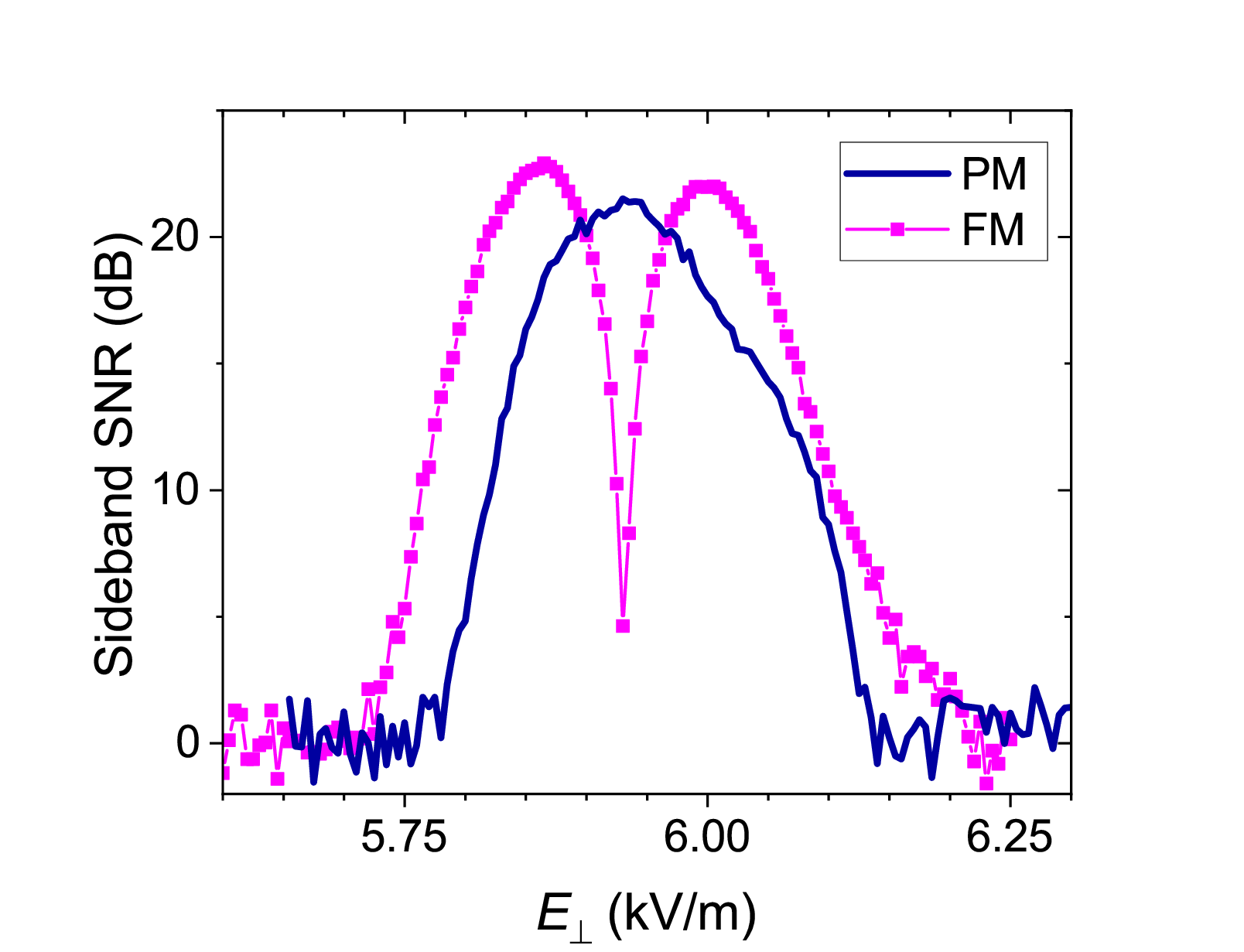}  
\caption{\label{fig:9} (color online) The sideband amplitude (in dB) versus the pressing electric field $E_\perp$ measured by applying PM (solid line) and FM (solid squares) mm-wave excitation at the frequency of 165~GHz.}
\end{figure} 

\begin{figure}[htp]
\includegraphics[width=\columnwidth,keepaspectratio]{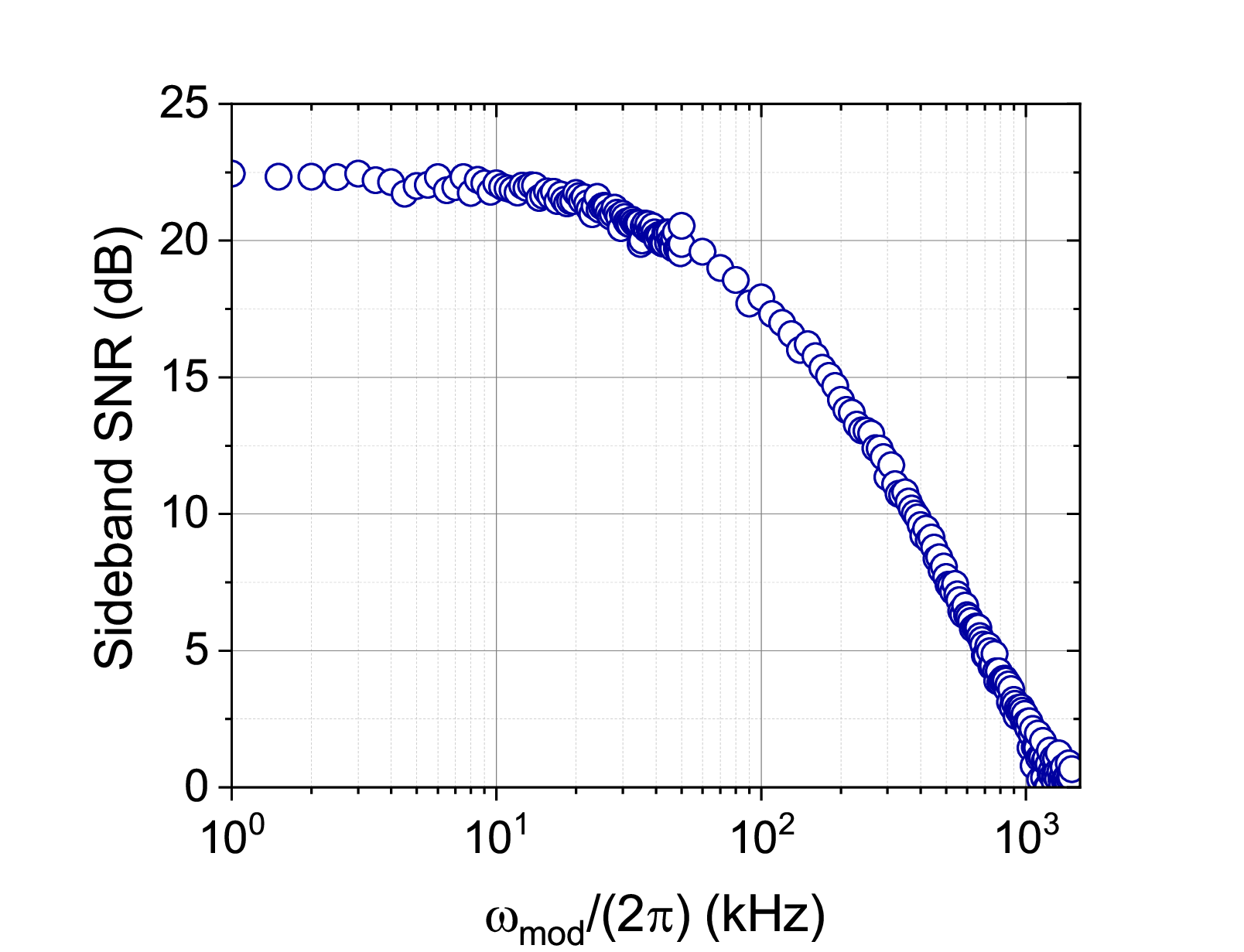}  
\caption{\label{fig:10} (color online) The sideband amplitude (in dB) versus the modulation frequency $\omm$.}
\end{figure}

We also investigated the dependence of the sideband signal on the modulation frequency $\omm$. Fig.~\ref{fig:10} shows the Bode plot of sideband SNR measured with electrons under the resonant mm-wave excitation. The dependance on the modulation frequency $\omm$ shows a characteristic low-pass-filter dependence with the cut-off (-3~dB) frequency of about 60~kHz. Such a high-frequency cut-off is not expected from the analysis of the electrical circuit shown in Fig.~\ref{fig:2}(b). Thus, it must be attributed to the time constant of a physical process in the mm-wave-excited electron system that causes the capacitive change in the setup impedance and the observed reflection response. Interestingly, the image-charge response is clearly observed at the modulation frequency exceeding 1~MHz, as shown in Fig.~\ref{fig:7}(b). This indicates that the physical mechanism of the observed reflection response is very different from the well-understood mechanism of the image-charge response. This will be discussed further in Sec.~\ref{sec:disc}.     

\section{Discussion and Conclusions}
\label{sec:disc}

In this work, we showed that the rf reflectometry method, which is based on the detection of small changes in the electrical impedance of a setup coupled to a many-electron system on liquid helium, serves as a very sensitive method for detecting the Rydberg resonance of such electrons. Remarkably, the sensitivity even exceeds that of the image-charge detection method employing a high-impedance $LCR$ circuit, which is successfully used for detection of harmonic motion and quantum states of a single trapped ion~\cite{winelandJAP1975,ulmPRL2011,haffPRL2017,winelandPRA2017hybrid}, and is also believed to be one of the promising routs towards detecting the Rydberg transition of a single electron on liquid helium~\cite{mishaPRA2025}. Therefore, it is hoped that rf reflectometry can be a new viable tool for quantum state readout in this system. However, the mechanism of the observed impedance change in response to the Rydberg-state excitation of electrons is not completely understood. As was originally conceived, the excitation of electrons from the ground state to the excited Rydberg state causes a state-dependent electric susceptibility of the system, therefore a capacitive contribution to the device impedance, which is proportional to the derivative of the excited-state population with respect to the transition frequency~\cite{kawakami2019image,kawakamiPRA2023qubits}. Thus, it is expected that for the proposed mechanism the rf reflection response vanishes for the maximum population of the excited state corresponding to the zero detuning of the mm-wave excitation from the Rydberg resonance. This result is confirmed by the theoretical calculation of the reflection response given in Sec.~\ref{sec:theory}. It is instructive to estimate the coupling strength $g$ of an electron to the resonant circuit and the expected reflection response in our setup. For a parallel-plate capacitor used in our setup, the coupling coefficient introduced in Sec.~\ref{sec:theory} is given by $g=e\Delta z V_0/(\hbar D)$, where $V_0=\sqrt{\hbar\omr/(2 C)}$ is the zero-point fluctuations of voltage in the circuit and $C \approx C_\textrm{par} + C_\textrm{p}$ is the circuit capacitance (see Fig.~\ref{fig:2}(b)). For our setup, we obtain $g\approx 250$~Hz. For the maximum Rabi frequency of mm-wave excitation on the order of 10~MHz~\cite{kawakamiPRL2021relax}, we obtain the dispersive shift of the resonator frequency on the order of  $\delta\omr = 10^{-4}$~Hz. For a single electron, the corresponding change in the reflection coefficient is minuscule. The response can be enhanced by coupling a large number of electrons ($N_\textrm{e}\sim 10^8$ ) to the resonator. In this case, one needs to take into account an inhomogeneous broadening of the transition line due to a non-uniform pressing electric field $E_\perp$. For the rf input power of -3~dBm used in this experiment and accounting for the attenuation, the mean number of photons in the resonator is expected to be on the order of $10^7$, which is much smaller than $\omc^2/(4g^2)$. It is straightforward to obtain an expression corresponding to Eq.~\ref{eq:Gamma} for an inhomogeneously broadened many-electron system in the form

\begin{equation}
\Gamma = -1 + \frac{\ke}{\frac{\kappa}{2}+i\Delta_0 - iN_\textrm{e}\delta\omr \tilde{\omega}_1 \int\limits_{-\infty}^{\infty} \frac{\epsilon' f(\epsilon - \epsilon')d\epsilon'}{\epsilon'^2 + (\gamma_{21}/2)^2 + \tilde{\omega}_1^2/2}},
\label{eq:Gamma2}
\end{equation}             

\noindent where $f(\epsilon)$ is a normalized distribution function characterizing the inhomogeneous broadening. The dispersive shift is proportional to $N_\textrm{e}$. However, there could be also an appreciable cancellation due to the convolution of $f(\epsilon)$ with an odd function of $\epsilon$. Most importantly, the reflection response retains the same dependence on the detuning $\epsilon$ as shown in Fig.~\ref{fig:1}, which is drastically different from what is seen in the experiment. It is clear that the rf response observed in the experiment must originate from a different mechanism.  

\begin{figure}[htp]
\includegraphics[width=\columnwidth,keepaspectratio]{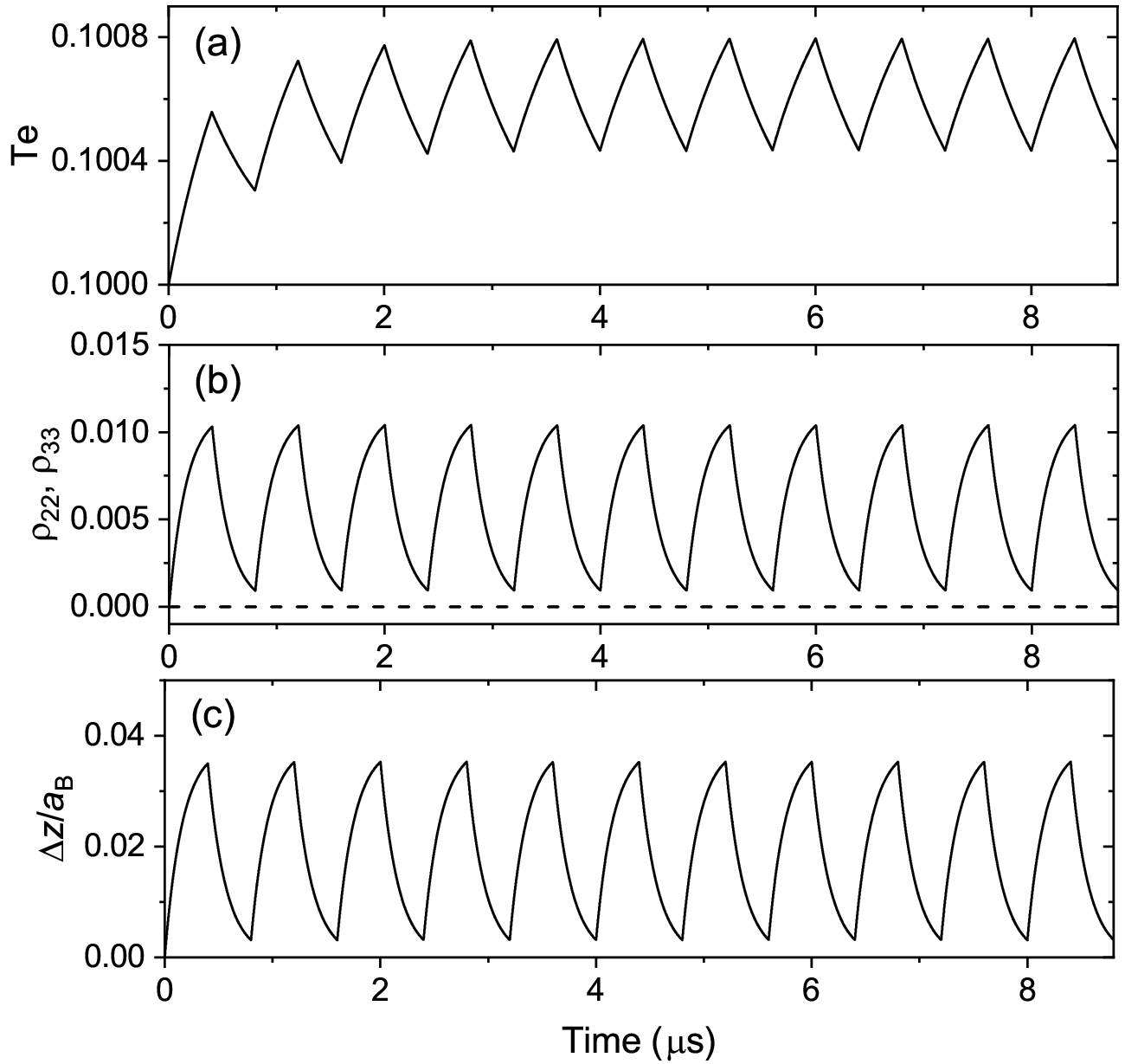}  
\caption{\label{fig:add} (color online) Time evolution of (a) the electron temperature $T_\textrm{e}$, (b) the first-excited (solid line) and the second-excited (dashed line) Rydberg state population, and (c) the ratio between the the mean vertical displacement of an excited electron $\Delta z$ and its effective Bohr radius $a_\textrm{B}$, calculated for an electron system under the pulse-modulated ($f_\textrm{mod}=1.2$~MHz) resonant mm-wave excitation with the Rabi frequency $\omega_1/2\pi = 300$~kHz at $T=0.1$~K.}
\end{figure}

Comparison between the reflection response due to, on the one hand, excitation of the Rydberg states of electrons and, on the other hand, modulation of their electrostatic confinement (see Figs.~\ref{fig:5}(a,b)) points out that the observed change in the device impedance could be due to in-plane motion of electrons induced by excitation of their Rydberg states. It is well known that electrons on liquid helium can be overheated by the mm-wave excitation to a temperature $T_\textrm{e}>T$ due to quasi-elastic scattering and decay of the excited electrons accompanied by transfer of the excitation energy into the kinetic energy of the lateral motion~\cite{saitJPSJ1978,konstPRL2007}. It is instructive to estimate the degree of the electron overheating under the condition of our experiment. Fortunately, it is possible to do it by analyzing the image-charge response obtained in Sec.~\ref{sec:image}. In the conventional picture, due to the fast electron-electron collisions accompanied by the exchange of energy between electrons, the system can be characterized by an effective electron temperature $T_\textrm{e}$, which is assumed to be the same for the entire electron system. An important point is that for sufficiently low $T$, when the intrinsic transition linewidth due to the scattering of electrons from the liquid helium excitations is less than the inhomogeneous broadening due to a nonuniform pressing electric field $E_\perp$, only a certain fraction $\alpha$ of electrons absorb energy from the radiation, while the entire system looses energy during inelastic scattering processes. For example, using an estimated transition linewidth of $\gamma_\textrm{hom}\sim 1$~MHz at $T=0.1$~K from the relaxation and dephasing rates obtained in Appendix~\ref{app:relax} and the inhomogeniously broadened transition linewidth of $\gamma_\textrm{inh} \sim 1$~GHz observed by the image-charge detection (see Fig.~\ref{fig:7}(b)), we expect that a fraction of only about $\gamma_\textrm{hom}/\gamma_\textrm{inh} \sim 0.1\%$ of all electrons gets excited by the applied radiation in our experiment. The electron temperature $T_\textrm{e}$ can be found from the energy balance equation per electron $\kB (dT_\textrm{e}/dt)\approx \alpha \hbar\ommm r(\rho_{11}-\rho_{22}) + \dot{E}$, where $\rho_{nn}$ is the fractional occupancy of the electrons that get excited by the radiation, $r$ is the excitation rate of electrons due to radiation, and $\dot{E}$ is the rate of the energy loss by electrons due to the inelastic scattering processes~\cite{kawakamiPRL2021relax}. The latter can be represented in a simplified form $\dot{E}=-\nu_E\times (T_\textrm{e}-T)$, where $\nu_E\sim 10^6$~s~\cite{monarkha2004}. This already allows for a quick estimation of the electron temperature in a steady state for a given excitation rate $r$. In the case of fast pulse modulation of the applied radiation ($f_\textrm{mod}\approx 1.2$~MHz in our experiment), with the modulation rate comparable to the energy relaxation rate, one needs to calculate the time evolution of the fractional occupancies $\rho_{nn}$ by solving the time-dependent rate equations~\cite{kawakamiPRL2021relax}. As an example, the calculated time evolution of the electron temperature $T_\textrm{e}$, the first-excited state and the second-excited state occupancies, $\rho_{22}$ and $\rho_{33}$, respectively, and the mean vertical displacement of an electron under irradiation $\Delta z $ are shown in Figs.~\ref{fig:add}(a), \ref{fig:add}(b), and \ref{fig:add}(c), respectively. To obtain this result, the coupled time-dependent energy balance equation and the rate equations for the occupancies $\rho_{nn}$ are numerically solved at $T=0.1$~K, as was previously described~\cite{kawakamiPRL2021relax}. In the calculations, we assume a resonant excitation rate $r=0.5\omega_1^2/\gamma_\textrm{hom}$, where the homogeneous Lorentzian half-width is given by $\gamma_\textrm{hom}=\gamma_{21}/2+\gamma_\phi$~\cite{CollinPRL2002}. The decay and dephasing rates $\gamma_{21}$ and $\gamma_\phi$ due to the elastic one-ripplon scattering are calculated using the expressions derived in Appendix~\ref{app:relax}. In principle, the two-ripplon scattering processes can also contribute to the decay rate. Most importantly, the inelastic two-ripplon emission process determines the energy relaxation rate $\dot{E}$, which depends on the state occupancies of all electrons~\cite{kawakamiPRL2021relax}. In the calculations, we use a small fraction $\alpha=0.1\%$ of the electrons which is excited by the radiation, therefore we assumed the thermal population corresponding to the electron temperature $T_\textrm{e}$ for the numerical evaluation of $\dot{E}$. In Fig.~\ref{fig:add}, the PM resonant radiation with the modulation period of 0.8~$\mu$s is turned on at $t=0$. It is seen that soon after it is turned on, the steady oscillations of both $T_\textrm{e}$ and $\rho_{nn}$ are established. The electron temperature stays very close to $T$, consequence of the fact that only a small fraction $\alpha$ of the electrons absorb energy from the applied radiation, while all electrons loose energy due to the inelastic scattering from ripplons. The population of the second-excited state (dashed line in Fig.~\ref{fig:add}(b)) and the higher excited states is negligible. Therefore, the mean vertical displacement of an excited electron can be calculated as $\Delta z = \rho_{22} (z_2 -z_1)$, where $z_n$ is the mean distance of an electron in the $n$-th Rydberg state. It is shown  in Fig.~\ref{fig:add}(c) in the units of the effective Bohr radius $a_\textrm{B}\approx 7.6$~nm. From Fig.~\ref{fig:add}(c), we can calculate the image charge induced at the top electrode of our setup as $Q=\alpha N_\textrm{e}(\Delta z/D)e$, where $N_\textrm{e}\approx 8.65\times 10^7$ is the total number of electrons comprising the surface charge (see Sec.~\ref{sec:demod}). We obtain $Q\approx 0.01 e$, which must be compared with our experimental estimate of $Q$ obtained from the image-charge detection in Sec.~\ref{sec:image}. We thus conclude that for the maximum excitation power used in our experiment the overheating of electrons is negligible.                             

From the results of calculations presented in Fig.~\ref{fig:add}(a), we must conclude that the overheating of electrons can not explain the rf response of the electron system to the mm-wave excitation observed in our experiment. However, the above calculations are based on the conventional assumption of the effective electron temperature, which is the same for the entire system. This assumption has been recently challenged by two independent observations of the thermoelectric transport of electrons in the system due to their local heating~\cite{lyonPRL2018,KostylevPRL2021}. In Ref.~\cite{KostylevPRL2021}, such a local heating of electrons was induced by selectively tuning a controlled fraction of the electron system into resonance with the applied mm-wave excitation, while their in-plane motion was detected by the image charge induced at the electrodes of a capacitor setup similar to that used in this work. We note that such a partial tuning of a fraction of about 0.1\% of the total number of electrons in the system is also realized in the experiment described here due to the inhomogeneous broadening, as discussed earlier. Thus, it is expected that the thermoelectric transport must be induced in the experiment described here, similar to the experiment in Ref.~\cite{KostylevPRL2021}. Unfortunately, inclusion of this effect into the Green's function calculations presents a rather challenging problem and is not attempted here. Moreover, the theoretical framework of the observed thermoelectric effect needs to be further investigated~\cite{KostylevPRL2021}. Nevertheless, it is worth to note that the direction of the in-plane transport of electrons in response to the mm-wave excitation is similar for the experiment in Ref.~\cite{KostylevPRL2021} and the experiment reported here. Indeed, comparison between the reflection responses shown in Fig.~\ref{fig:5}(a) and Fig.~\ref{fig:5}(b) indicates that electrons flow from the center of the charged layer towards its edge when mm-wave excitation is switched on (in Fig.~\ref{fig:5}(a) this corresponds to the situation when the voltage applied to the guard electrodes $V_\textrm{BGTG}$ becomes more positive). Similar behaviour was observed in the experiment of Ref.~\cite{KostylevPRL2021}. The possibility of the thermoelectric transport and its role in the experiment reported here requires further investigation.

In conclusion, the Rydberg resonance of microwave-excited electrons on the surface of liquid helium is observed by the rf reflectometry method as the variation of the electrical impedance of the system. While mechanism of the impedance change is not entirely understood, it is likely associated with the in-plane transport of electrons in response to the resonant excitation. Remarkably, the method demonstrates an unprecedented sensitivity exceeding that of the image-charge detection. This shows that the rf reflectometry could be a viable experimental technique to study the interesting many-electron dynamics of photo-excited electrons, where some surprising collective phenomena has been observed~\cite{zudovPRL2003,maniNature2022,dorozhNatPhys2011,konstPRL2010,chepelNatCommun2015}. Regarding the prospects for the quantum state detection, it is clear that one needs to enhance the coupling constant $g$. The simplest solution is to increase the ratio $\Delta z/D$. It was already demonstrated that this can be done by confining electrons in a superfluid-filled microchannel, which allowed to increase this ratio by approximately thousandfold~\cite{zouNJP2022image}. Towards a single electron detection, an enhancement in the rms vacuum voltage $V_0$ by increasing the resonant circuit impedance could be crucial. The coplanar waveguide (CPW) resonators with $\omr/(2\pi)\sim 5$~GHz fabricated from high-kinetic inductance material proved to be an excellent choice of a resonant circuit for strong coupling~\cite{koolstraPRA2025}. However, the dispersive shift in our case scales as an inverse square of the resonant frequency $\omr$, therefore a resonant circuit with a lower frequency is desired. The lumped-element circuits employing superconducting spiral inductors allow a wide-range of resonant frequencies and could be preferred~\cite{blokPRA2025}.      
                 
{\bf Acknowledgements} We acknowledge helpful discussions with Dr. Erika Kawakami and the members of her group. This work is supported by the internal grant from the Okinawa Institute of Science and Technology (OIST) Graduate University and the Grant-in-Aid for Scientific Research (Grant No. 23H01795 and 23K26488) KAKENHI MEXT.

$   $
\appendix 

\section{Derivation of the relaxation and dephasing rates}
\label{app:relax}

We start by considering an electron bound to the surface of liquid helium and coupled to the surface ripplons. It is assumed that the in-plane motion of electron is not quantized. The Hamiltonian of the system can be written as

\begin{equation}
H = K_\textrm{e} + \sum\limits_{n>1} \hbar\omega_{n1}|n\rla n| + \sum\limits_\vq \hbar\omega_q b_\vq^\dagger b_\vq + H_\textrm{e-r},
\label{eq:Hfull}
\end{equation}  

\noindent where $K_\textrm{e}$ is the electron kinetic energy of the in-plane motion, $\omega_{n1}$ corresponds to the energy difference between the ground Rydberg state and the Rydberg states with the quantum number $n$, $b_\vq^\dagger$ ($b_\vq$) is the bosonic creation (annihilation) operator for ripplons with the wave vector $\vq$, and the dispersion relation of ripplons is given by $\omega_q = \sqrt{\alpha/\rho}q^{3/2}$ ($\alpha=0.37\times 10^{-3}$~J/m$^2$ and $\rho=0.145\times 10^3$~kg/m$^3$ is the surface tension and density of liquid helium, respectively). The electron-coupling Hamiltonian is given by  

\begin{equation}
H_\textrm{e-r} = \sum\limits_\vq Q_q U_q(z) e^{i\vq\vr} \left( b_\vq + b_{-\vq}^\dagger \right),
\label{eq:Her}
\end{equation}  

\noindent where $Q_q=\sqrt{\hbar q/(2S\rho\omega_q)}$ ($S$ is the surface area occupied by electron) and the electron-ripplon coupling $U_q$ can be represented as

\begin{equation}
U_q(z) = -\frac{dV_\textrm{e}}{dz} + eE_\perp + \frac{e^2q^2(\epsilon_\textrm{He} -1)}{16\pi\epsilon_0(\epsilon_\textrm{He} +1)} \left[ \frac{1}{(qz)^2} - \frac{K_1(qz)}{qz} \right],
\label{eq:Uq}
\end{equation}  

\noindent where $\epsilon_0$ is the electric permittivity of vacuum, $\epsilon_\textrm{He}$ is the dielectric constant of liquid helium, and $V_\textrm{e}(z)$ is the potential energy of an electron at the flat interface~\cite{monarkha2004}. It is convenient to represents the electron-dependent part of \eqref{eq:Her} as an expansion over the electron states $|\vk n\rangle$, where $\vk$ is the in-plane wave vector of electron, as  

\begin{equation}
U_q(z)e^{i\vq\vr} = \sum\limits_{\vk,n}\sum\limits_{\vk',n'} \left( U_q(z)e^{i\vq\vr} \right)_{\vk n,\vk' n'} |\vk n\rangle\langle \vk' n'|.
\label{eq:Uqexp}
\end{equation}

\noindent Then, the coupling Hamiltonian \eqref{eq:Her} in the interaction picture can be written as

\begin{eqnarray}
&& H_I = \sum\limits_{n,n'} \sum\limits_{\vq,\vk} Q_q (U_q)_{nn'} |\vk n\rla (\vk -\vq) n'| e^{i(\omega_\vk - \omega_{\vk-\vq} + \omega_{nn'})t} \nonumber \\
&& \times (a_\vq e^{-i\omega_qt} + a_{-\vq}^\dagger e^{i\omega_qt} ),
\label{eq:HI1}
\end{eqnarray}

\noindent where $(U_q)_{nn'}$ is the matrix element of $U_q$ with respect to the Rydberg states $|n\rangle$ and $|n'\rangle$, and $\omega_\vk = \hbar k^2/(2\me)$ ($\me$ is the electron mass). 

We use the master equation for the electron density operator $\rho$, which in the Born-Markov approximation can be represented as~\cite{Carmichael}

\begin{equation}
\frac{d\rho}{dt} = -\frac{1}{\hbar^2}\int\limits_{-\infty}^t\tr \big\{[H_I(t),[H_I(t'),\rho(t)\rho_\textrm{R}]]\big\}dt',
\label{eq:master1}
\end{equation}

\noindent where trace is over the states of the ripplon system in thermal equilibrium at temperature $T$. In what follows, we neglect population of the Rydberg states with $n>2$ and represent the Hamiltonian \eqref{eq:HI1} in the form of a standard spin-boson model as

\begin{equation}
H_I = \sigma_z (F_1 + F_1^\dagger) + \left( \sigma_+ e^{i\omega_{21}t} + \sigma_- e^{-i\omega_{21}t} \right) (F_2 + F_2^\dagger),
\label{eq:HI2}
\end{equation} 

\noindent with the fluctuating field operators given by

\begin{eqnarray}
&&  F_1 (t) = \frac{1}{2}\sum\limits_{\vq,\vk} Q_q \left[ (U_q)_{22} - (U_q)_{11})\right] |\vk\rangle\langle \vk -\vq| a_\vq e^{-i(\omega_q - \omega_\vk + \omega_{\vk-\vq})t}, \nonumber \\ 
&&  F_2 (t) = \sum\limits_{\vq,\vk} Q_q (U_q)_{21} |\vk\rangle\langle \vk -\vq| a_\vq e^{-i(\omega_q - \omega_\vk + \omega_{\vk-\vq})t}.
\label{eq:F}
\end{eqnarray}

\noindent The first term in \eqref{eq:HI2} leads to dephasing, while the second term leads to the relaxation between the Rydberg states. We can consider these two process separately due to a large difference in the time dependance owing to the factor $ e^{\pm  i\omega_{21}t}$. Substituting the Hamiltonian \eqref{eq:HI2} into \eqref{eq:master1}, tracing over the states of the ripplon bath, and taking the integral over time using the well-known representation

\begin{equation}
\int\limits_0^{+\infty} e^{\pm i(\omega - \omega_0) \tau} d\tau = \pi\delta(\omega-\omega_0) \pm i \frac{P}{\omega - \omega_0},
\label{eq:integral}
\end{equation} 

\noindent where $P$ stands for the Cauchy principle value, we obtain

\begin{widetext} 

\begin{eqnarray}
&&  \frac{d\rho}{dt} = -\frac{\pi}{2\hbar^2} \Big[ \sum\limits_{\vq,\vk} Q_q^2 \left[(U_q)_{22} - (U_q)_{11}\right]^2 (2\bar{n}_q+1) |\vk\rangle\langle\vk| \delta (\omega_\vk - \omega_{\vk-\vq}) \Big] (\rho  - \sigma_z\rho\sigma_z ) \nonumber \\ 
&&  - \frac{\pi}{\hbar^2} \Big[ \sum\limits_{\vq,\vk} Q_q^2 (U_q)_{21}^2 (2\bar{n}_q+1) |\vk\rangle\langle\vk| \delta (\omega_\vk - \omega_{\vk-\vq} +\omega_{21}) \Big] \left(\{\sigma_+\sigma_-,\rho\} - 2\sigma_-\rho\sigma_+ \right) \nonumber \\
&&  - \frac{\pi}{\hbar^2} \Big[\sum\limits_{\vq,\vk} Q_q^2 (U_q)_{21}^2 (2\bar{n}_q+1) |\vk\rangle\langle\vk| \delta (\omega_\vk - \omega_{\vk-\vq} -\omega_{21}) \Big] \left(\{\sigma_-\sigma_+,\rho\} - 2\sigma_+\rho\sigma_- \right).
\label{eq:master2}
\end{eqnarray}

\end{widetext} 

\noindent For the sake of brevity, we omitted the terms coming from the second part on the left-hand-side of Eq.~\eqref{eq:integral}, which result in the Lamb-like shifts of the electron energy levels. Also, we neglected the ripplon energy in the delta function, which for quasi-elastic one-ripplon scattering processes is much smaller than the electron energy~\cite{monarkha2004}. Finally, tracing over the states of the electron in-plane motion at the thermal equilibrium at temperature $\Te$ according to $\sum\limits_\vk \langle\vk|...|\vk\rangle f_\vk$, with the Gibbs distribution $f_\vk\propto \exp[-\hbar^2 k^2/(2m_\textrm{e}k_B\Te)]$, we obtain the master equation for the reduced density operator $\rho_\textrm{s}$ in the form of Eq.~\eqref{eq:master} in the main text, with the dephasing and relaxation rates given by

\begin{eqnarray}
&&  \gamma_\phi = \frac{T\sqrt{k_B}}{8\sqrt{\pi T_e}\alpha\hbar} \int\limits_0^{\infty} \frac{dE_q}{E_q^{3/2}} \left[(U_q)_{22} - (U_q)_{11}\right]^2 e^{-\frac{E_q}{4 k_BT_e}}, \nonumber \\ 
&&  \gamma_{nn'} = \frac{T\sqrt{k_B}}{4\sqrt{\pi T_e}\alpha\hbar} \int\limits_0^{\infty} \frac{dE_q}{E_q^{3/2}} \left[(U_q)_{21}\right]^2 e^{-\frac{(E_q-\hbar\omega_{nn'})^2}{4E_q k_BT_e}},
\label{eq:rates}
\end{eqnarray}

\noindent where $E_q=\hbar^2q^2/(2m_\textrm{e})$ and $n,n'=1,2$ ($n\neq n'$). It is clear that $\gamma_{12}=\gamma_{21}e^{-\hbar\omega_{21}/(2m_\textrm{e}k_B\Te)}$, as should be expected from the principle of detailed balance. For typical energy difference $\hbar\omega_{21}/k_B\sim 8$~K and electron temperature $\Te\lesssim 1$~K, we have $\gamma_{12}<< \gamma_{21}$, and the last term in master equation Eq.~\eqref{eq:master} in the main text can be safely neglected. For the pressing field $E_\perp = 6$~kV/m and temperature $T=0.1$~K used in the experiment, we calculate the relaxation rate $\gamma_{21}/(2\pi)=1.3$~MHz. This is comparable to the decay rate by a spontaneous two-ripplon emission~\cite{monarkhaJLTP2007}. Under the same conditions, the pure dephasing rate is $\gamma_\phi/(2\pi) = 64$~kHz, thus it can be neglected comparing to $\gamma_{21}/2$.     

\section{Calculation of the cell impedance}
\label{app:green}

\begin{figure}[htp]
\includegraphics[width=\columnwidth,keepaspectratio]{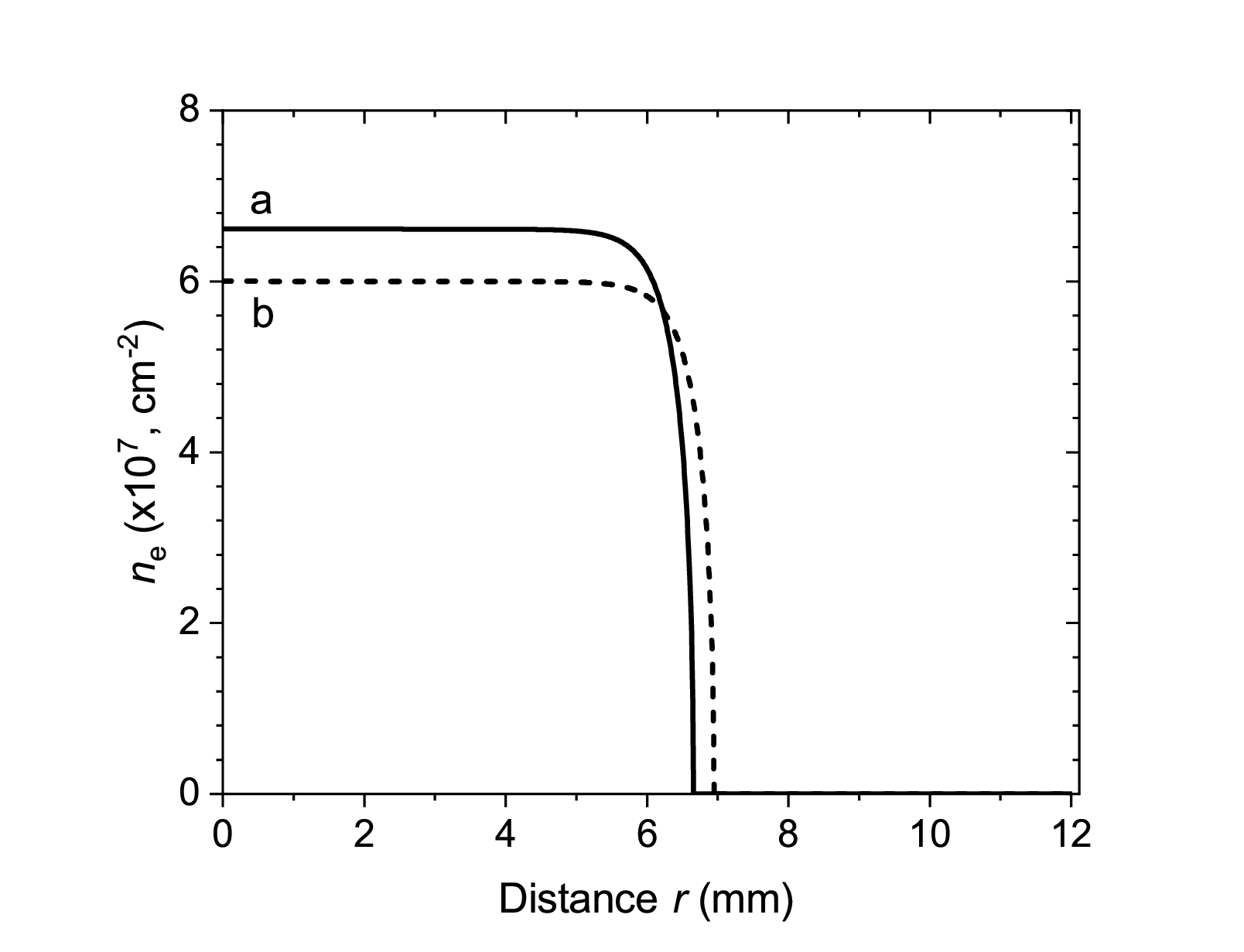}  
\caption{\label{fig:11} (color online) The distribution of areal density of surface electrons $n_\textrm{e}$ for two different sets of bias voltages applied to the electrodes (a) $V_\textrm{BCBM}=30$~V, $V_\textrm{TC}=V_\textrm{TM}=18$~V, $V_\textrm{BG}=V_\textrm{TG}=-60$~V and (b) $V_\textrm{BCBM}=30$~V, $V_\textrm{TC}=V_\textrm{TM}=18$~V, $V_\textrm{BG}=V_\textrm{TG}=-25$~V. Note that the distribution (a) corresponds to the complete screening of the electric field above the charged surface.} 
\end{figure}

The Green's function method is used to calculate the electrical impedance of the cell containing the electron system~\cite{wilen1988jltp}. We consider a cylindrical cell of radius $R=1.8$~cm and height $H=0.2$~cm. The distribution of the electrostatic potential $\phi(\textbf{r})$, with $\textbf{r}=(r,z)$, and the electron density $n_\textrm{s}(r)$ satisfy the integral equation

\begin{equation}  
\phi(\textbf{r}) = \tilde{\phi}(\textbf{r}) + 2\pi\int G(\textbf{r},r') n_\textrm{s}(r') dr',
\label{eq:int}
\end{equation}       

\noindent where $\tilde{\phi}$ is the electrostatic potential due to the bias voltages applied to the concentric electrodes at the bottom and top of the cell and $G(\textbf{r},r')$ is the Green function corresponding to the potential at a point with coordinates $\textbf{r}$ due to a ring of charge with unit charge density located at radius $r'$ on the surface of liquid. To find the density distribution, Eq.~\eqref{eq:int} is solved by the finite difference method (FDM) on a 2D coordinate grid of dimensions $1800\times 200$. The surface of liquid is assumed to be located at the middle of the cell ($z=H/2$), and the dielectric constant of liquid helium is assumed to be unity for simplicity. Following Ref.~\cite{wilen1988jltp}, the potential $\tilde{\phi}(\textbf{r})$ and the Green functions $G(\textbf{r},r')$ are found by the relaxation method, and the density distribution at electrostatic equilibrium is found by assuming vanishing potential difference (electric field) within the charged surface. The equilibrium density profiles calculated for two different sets of bias voltages applied to the electrodes in our experiment are shown in Fig.~\ref{fig:11}.

\begin{figure}[htp]
\includegraphics[width=\columnwidth,keepaspectratio]{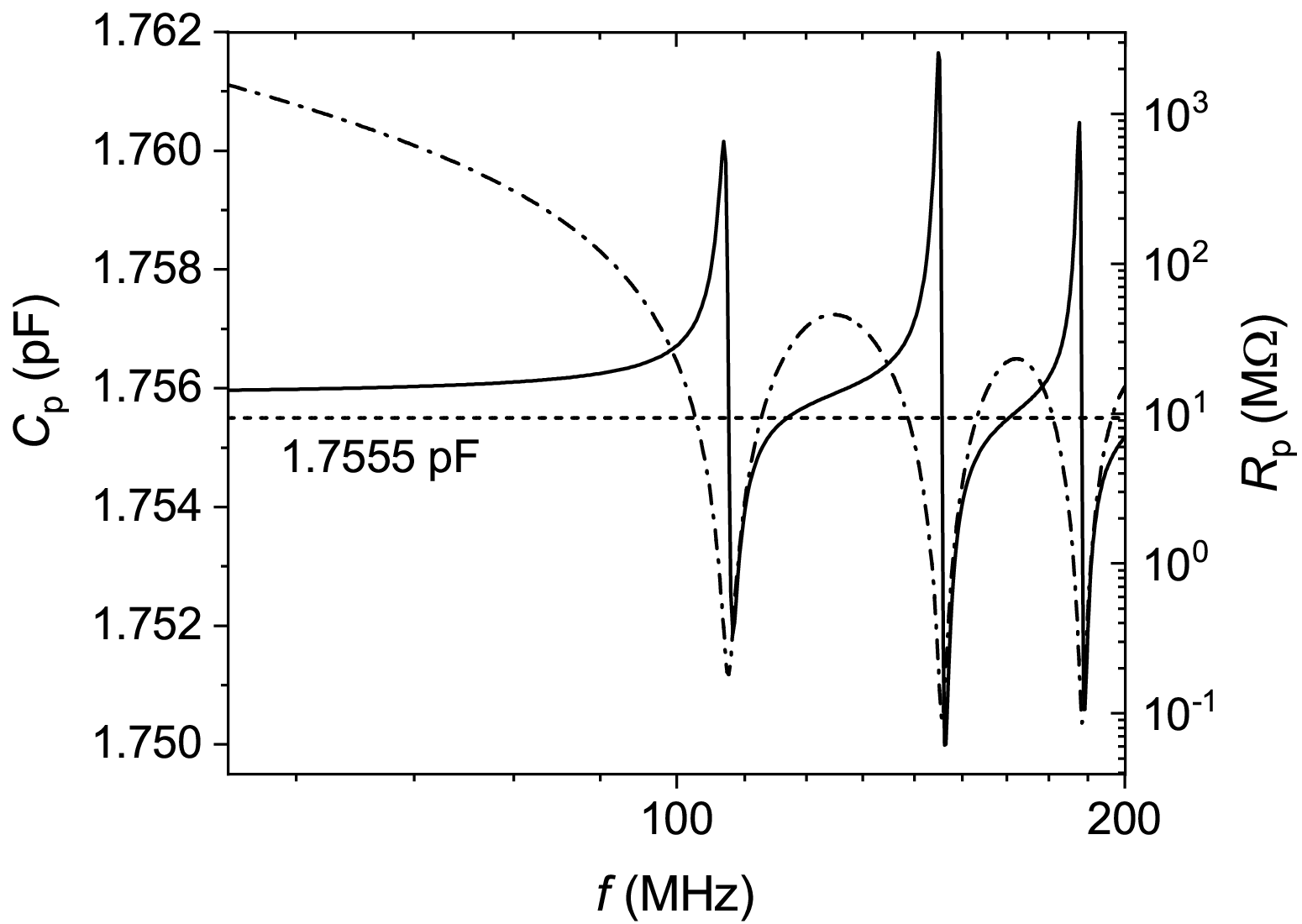}  
\caption{\label{fig:12} (color online) Capacitive (solid line, vertical axis on the left) and resistive (dash-dotted line, vertical axis on the right) components of the cell impedance as a function of the driving frequency $f=\omega/(2\pi)$ calculated for the electron system under the applied bias voltages corresponding to the equilibrium density profile given by the dashed line in Fig.~\ref{fig:11}. The dashed line indicates the calculated value of capacitance $C_\textrm{p}$ of an empty cell.} 
\end{figure}

The electrical impedance of the cell is defined as $Z_\textrm{cell}=V/I$, where $V$ is the amplitude of the rf driving voltage applied to the detection electrode and $I$ is the current flowing into this electrode~\cite{wilen1988jltp}. For a given equilibrium density profile, we assume small variation of the density $\delta n_\textrm{s}(\textbf{r})$ due to a small voltage signal $V=20$~mV and calculate the distribution of the corresponding electric potential $\delta\phi(\textrm{r})$ in the cell by taking into account the linearized continuity equation at the charged surface with the current density $j(r)=-\sigma \partial (\delta \phi)/\partial r$, where $\sigma$ is the electrical conductivity. An expression for the electric conductivity can be found from the general balance-of-force equation as 

\begin{equation}
\sigma = \frac{e^2 n_\textrm{e}}{m_\textrm{e}}\frac{1}{\left[ \nu_\textrm{eff}(\omega)+i (\omega + w(\omega))\right]},
\label{eq:sigma}
\end{equation}

\noindent where the effective collision frequency $\nu_\textrm{eff}$ and the medium response $w$ can be also identified as the imaginary and real parts, respectively, of the memory function~\cite{monarkha2004}. In general, they depend on frequency and electron density. For high driving frequency $\omega >> \omega_{\textrm{g}_1}$, where $\omega_{\textrm{g}_1}$ is the ripplon frequency corresponding to the first reciprocal lattice vector of WS, the medium response results in a gaped optical branch of plasmon excitation of electrons. In our calculations, the medium response is represented by $\omg^2/\omega$, which reproduces a gaped plasmon spectrum, with $\omg$ used as an adjustable parameter to reproduce the experimetal data. Moreover, it was shown that in a large range of parameters, such as temperature and electron density, the collision rate that enters ac conductivity is very close to the momentum relaxation rate $\nu$ of the non-degenerate electron gas~\cite{monarkha2004,monarkhaJPSJ2001}. Therefore, in our calculation we used an expression for the ac conductivity given by Eq.~\eqref{eq:cond} in the main text. The current to the detection electrode is found from the calculated change in the induced charge at the detection electrode $\Delta Q$ by the relation $I=i\omega \delta Q$. Using the equivalent representation in Fig.~\ref{fig:1}(b), the calculated impedance is given by $Z(\omega)= \left[ i\omega C_\textrm{p}(\omega) + R_\textrm{p}^{-1}(\omega) \right]^{-1}$. An example of calculated $C_\textrm{p}$ and $R_\textrm{p}$ versus the driving frequency $\omega$ is shown in Fig.~\ref{fig:12}. Here, we used an equilibrium distribution of the electron density $n_\textrm{s}$ given by the dashed line in Fig.~\ref{fig:11}. This corresponds to the guard voltage setting marked by vertical dashed line in Figs.~\ref{fig:5}(a,b) in the main text. In Fig.~\ref{fig:12}, the three lowest resonant modes of the plasmon excitation are clearly seen. In the calculations, we used frequency $\omf$ as an adjustable parameter such that the first resonant plasmon mode coincides with the resonant frequency of the tank circuit of 108.4~MHz. This reproduces the observed plasmon resonance shown in Fig.~\ref{fig:6} in the main text.    

\noindent 
    
\section{Calculation of the reflection response}
\label{app:reflection}
 
\begin{figure}[htp]
\includegraphics[width=\columnwidth,keepaspectratio]{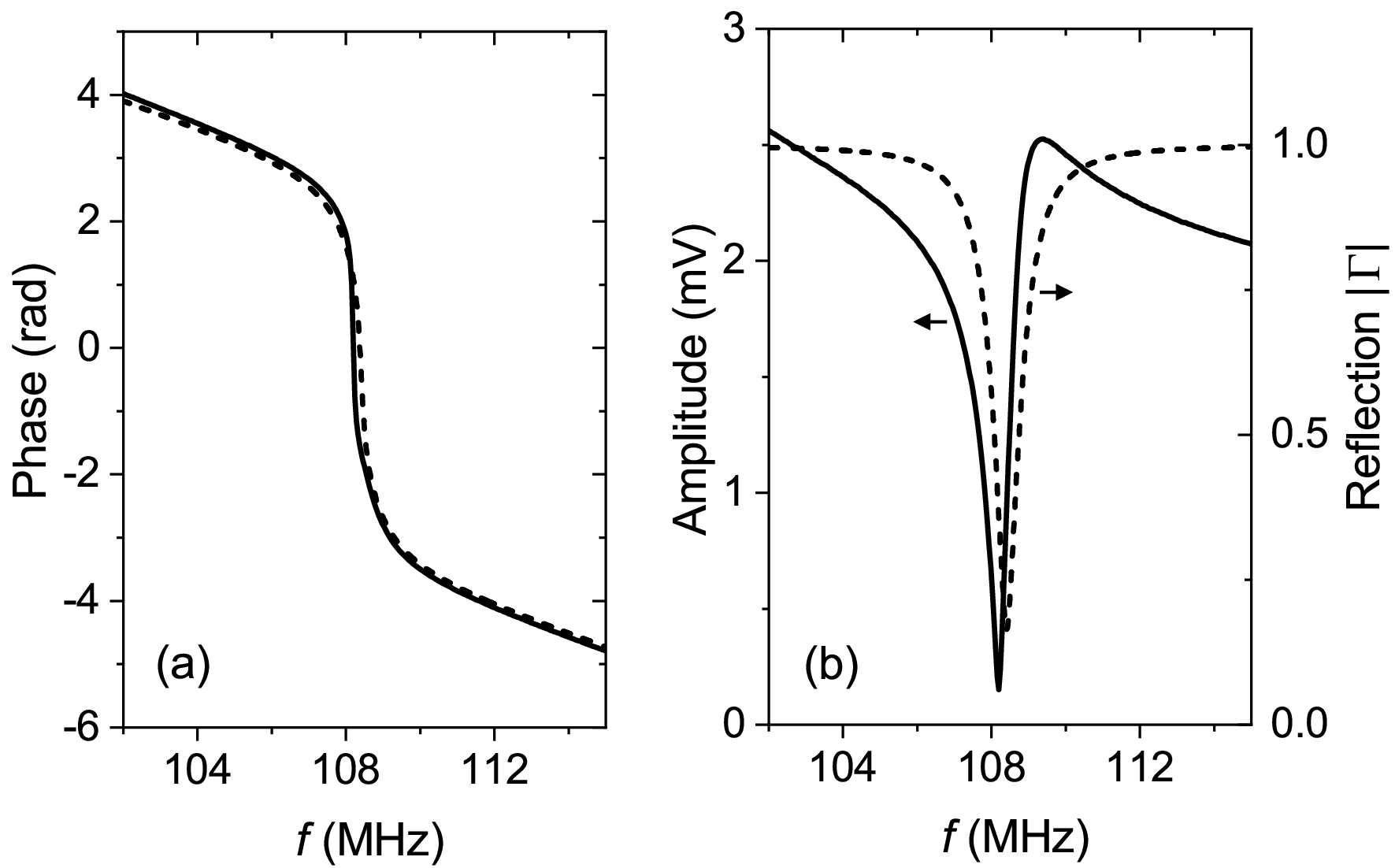}  
\caption{\label{fig:13} (color online) Phase (a) and amplitude (b) of the reflection signal (solid lines) measured with a lock-in amplifier, as described in Sec.~\ref{sec:setup}, and calculated (dashed lines) using the circuit model, as described in the text.} 
\end{figure}

\begin{figure}[htp]
\includegraphics[width=\columnwidth,keepaspectratio]{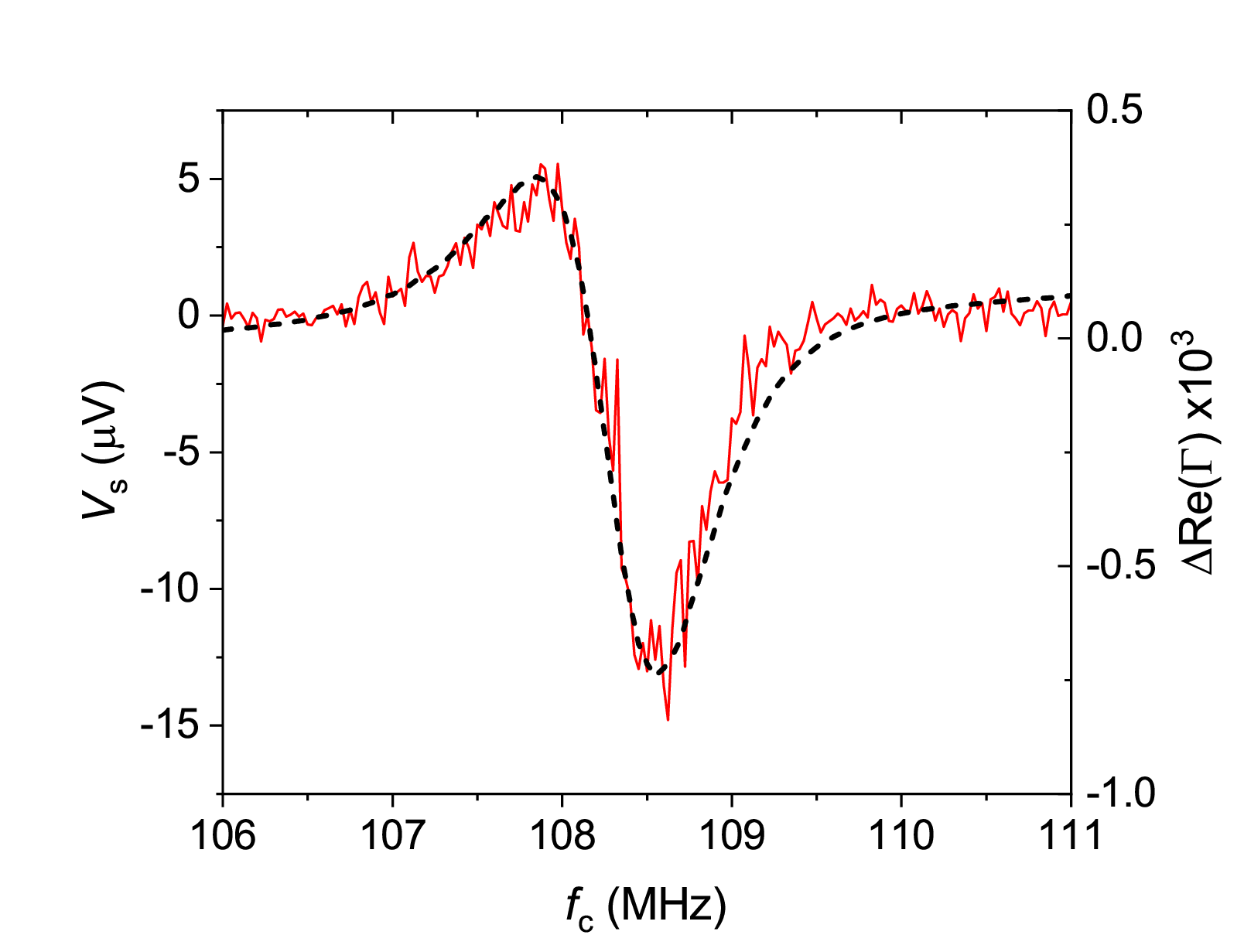}  
\caption{\label{fig:14} (color online) Demodulated reflection signal (solid line) measured by applying ac voltage modulation of 3~V$_\textrm{p-p}$ to the guard electrodes, with the dc voltage settings corresponding to $V_\textrm{BGTG}=-60$~V in Fig.~\ref{fig:5}(a) in the main text. The dashed line is the simulated response using the model circuit parameters described in the text.} 
\end{figure}

In order to make comparison with the observed rf reflection response, we calculated the reflection coefficient $\Gamma$ for the circuit model shown in Fig.~\ref{fig:2}(b) in the main text. We assumed that the reflection coefficient is given by

\begin{equation}
\Gamma = \frac{Z(\omega) - Z_0}{Z(\omega) + Z_0},
\label{eq:G}
\end{equation}

\noindent where $Z_0=50$~$\Omega$ is the impedance of the transmission line and $Z$ is the total impedance from the circuit's input port. In order to account for a frequency-dependent accumulation of phase in the transmission line connection between the circuit's input port and the room-temperature detector, we corrected the phase of the calculated reflection $\Gamma$ by adding an empirical phase shift (in radian) $-0.192 f [\textrm{MHz}] + 19.83$ obtained from the reflection spectrum measured as described in Sec.~\ref{sec:setup}. Fig.~\ref{fig:13} shows the comparison between the measured (solid lines) phase (a) and amplitude (b) of the reflected signal and the corresponding quantities (dashed lines) calculated from the circuit model with $L=0.777$~nH,  $C_\textrm{par}=1.645$~pF, $R_\textrm{L}=0.5$~$\Omega$, $C_\textrm{L}=0.3$~pF, $R=2.5$~$\Omega$, $C_1=10$~pF and $C_2=95$~pF. Note that the voltage-tunable varactor was added only in later experiments, therefore is not considered in this model. In this calculation, we assumed the cell impedance without electrons corresponding to $C_\textrm{p}=1.7555$~pF and $R_\textrm{p}=\infty$. Together with the chosen value of the parasitic capacitance $C_\textrm{par}$, the above value of $C_\textrm{p}$ determines the resonant frequency of the reflection signal of 108.43~MHz, which is close to the one independently obtained from the fitting described in Sec.~\ref{sec:setup}. Note that, unlike the fitting method used in Sec.~\ref{sec:setup}, our circuit model does not account for the asymmetry of the signal by an arbitrary impedance element representing impedance mismatch between a feed line and resonator~\cite{prob2015rsi}. Also, we found that the above choice of $C_2$, which differs from the capacitance (56~pF) of a surface-mount capacitor used in the circuit, gives a better fit to the quality factor and coupling that match those obtained from the fitting described in Sec.~\ref{sec:setup}. A likely reason for this deviation is that the input port of the tank circuit PCB is connected to the transmission line through the directional coupler (see Fig.~\ref{fig:2}(b) in the main text), which effectively reduces its coupling to the feedline.  Using this model, we can find the change in the reflection signal due to the variation in the cell impedance and compare it with the experimental results. As an example, Fig.~\ref{fig:14} shows comparison between the measured demodulated refection response (solid line) and the simulated response (dashed line). The measured response was obtained by applying 3~V$_\textrm{p-p}$ modulation in addition to a dc bias of -60~V to the guard electrodes (see Fig.~\ref{fig:5}(a) in the main text). Correspondingly, the simulated response is given by the difference between the in-phase components of the reflection coefficients calculated for two values of the guard voltage of $-58.5$ and $-60$~V. The agreement is excellent. From this calculations, we find that the above voltage modulation causes the variations of the capacitive and resistive contributions to the cell's impedance of about $\pm 40$~pF and $\mp 0.17$~M$\Omega$, respectively, with a corresponding variation of the electron pool radius of $\pm 0.02$~mm. The results of simulations for different dc bias voltages applied to the guard electrodes are shown in Fig.~\ref{fig:6} in the main text.

\section{Stark shift and peak structure}
\label{app:stark}

\begin{figure}[htp]
\includegraphics[width=\columnwidth,keepaspectratio]{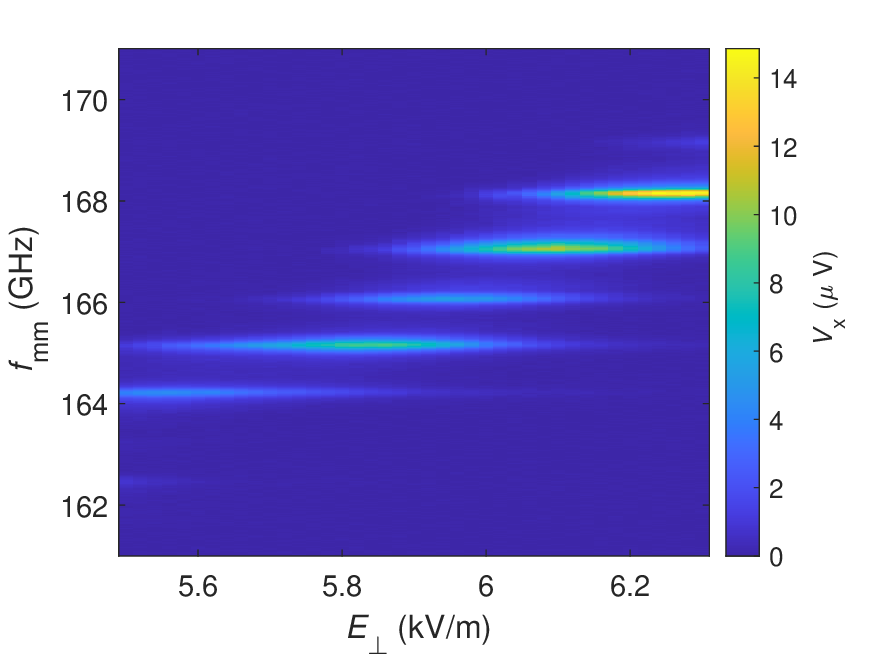}  
\caption{\label{fig:15} (color online) Color map of the demodulated reflection signal versus the pressing field $E_\perp$ and mm-wave frequency $f_\textrm{mm}$ measured with the lowest incident mm-wave power corresponding to the maximum attenuation in Fig.~\ref{fig:8} in the main text.} 
\end{figure}

The Rydberg transition frequency of electrons can be varied via the Stark shift by varying the value of the perpendicular pressing field $E_\perp$ exerted on the electrons. Fig.~\ref{fig:15} shows a color map of the reflection response versus $E_\perp$ and the mm-wave frequency $f_\textrm{mm}$ measured for the lowest excitation power used to obtain data shown in Fig.~\ref{fig:8} in the main text. It is clear that, while the Rydberg transition frequency of electrons varies with $E_\perp$, the frequency position of the observed signal peaks does not change. This points out that these peaks originate from the properties of the experimental setup, rather than the electron system. It is reasonable to suggest that such enhancements of signal, which appear at the discrete equidistant values of the radiation frequency, originate from the formation of standing waves of the mm-wave field due to multiple reflection of the incident radiation from the inner walls of the cell. Note that the vertical component of the radiation electric field is required to excite the Rydberg transition of electrons. If we assume an azimuthally symmetric TM mode of the radiation field inside the cell, for which the variation of the vertical electric field $E_z$ with the distance from the center of the cell $r$ is proportional to $J_0(\beta_{0(m)}r/R)$, where $J_0$ is the zero-order Bessel function of first kind, $\beta_{0(m)}$ is its $m$-th zero, and $R=20$~mm is the inner radius of the cell. Assuming $\beta_{0(m)}/R\approx \ommm/c$, where $c$ is the speed of light, we estimate the mode number $m=22$ for the mm-wave frequency of 166~GHz. For such a high mode number, the Bessel function is proportional to $\cos(\beta_{0(m)}r/R - \pi/4)$, from which we can estimate the frequency difference between adjacent resonant modes as $c/(2R)=7.5$~GHz. This is significantly larger than the observed frequency separation of about 1~GHz between the signal maxima in Fig.~\ref{fig:8}. However, it is likely that the actual distribution of the mm-wave field inside the cell, with the cross-section shown in Fig.~\ref{fig:2}(a), is more complicated than a single-mode field considered above. The numerical calculation of such field distribution is rather complicated and is not considered here.

\section{Sideband detection and capacitance sensitivity}
\label{app:sideband}

\begin{figure}[htp]
\includegraphics[width=\columnwidth,keepaspectratio]{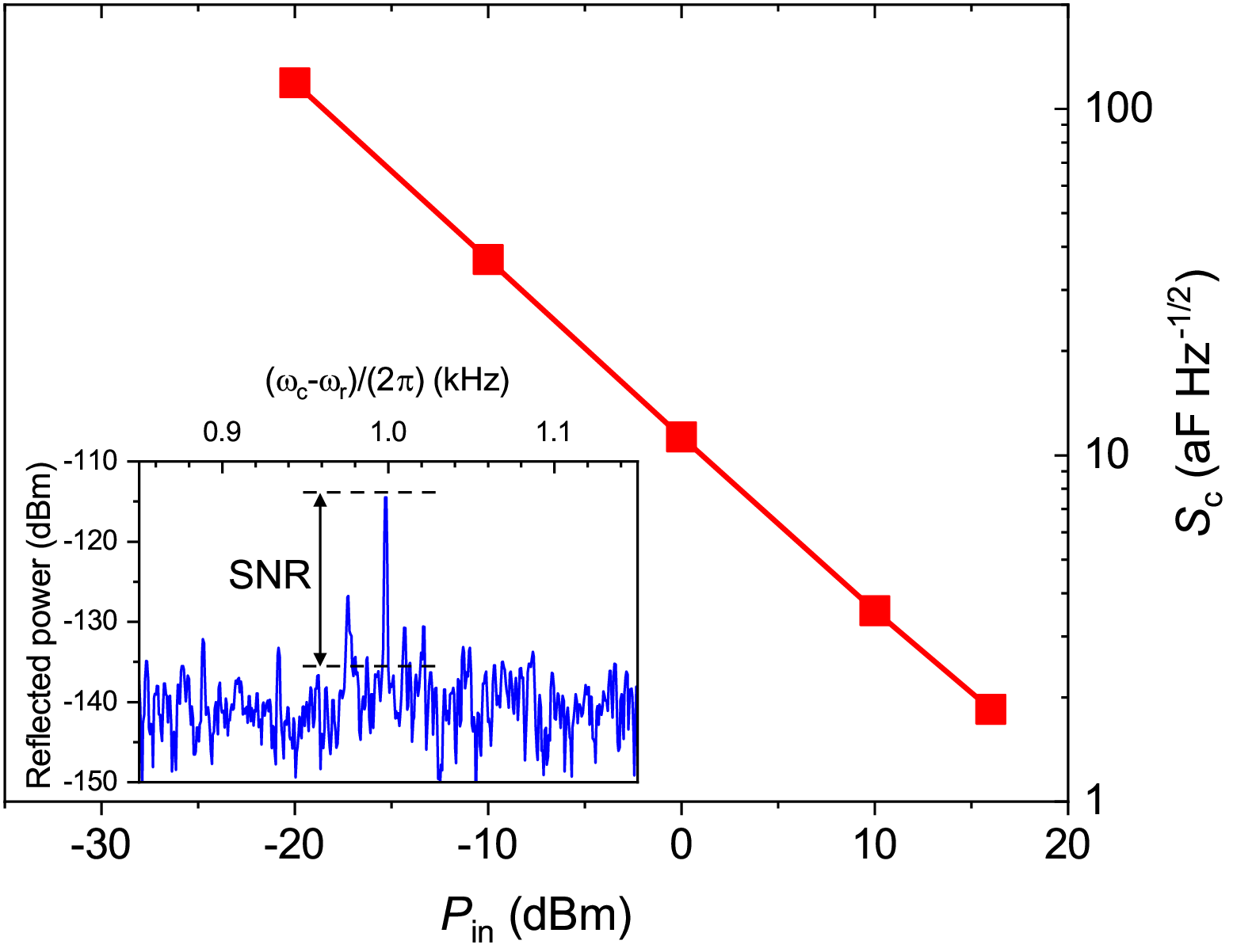}  
\caption{\label{fig:16} (color online) Sensitivity of the detection method to capacitive changes in the setup impedance obtained by applying an ac modulating voltage at the frequency $\omm/(2\pi)=1.2$~kHz to a voltage-tunable varactor integrated into the electrical device versus the input rf power $P_\textrm{in}$. In the inset: an example of the reflection power spectrum measured for electrons under PM ($\omm/(2\pi)=1$~kHz) mm-wave excitation.}
\end{figure}

Similar to Ref.~\cite{kawakamiPRL2025}, the rf reflection response of the system to the modulated Rydberg excitation is detected by appearance of a sideband in the reflection power spectrum. An example of power spectrum measured by SA is shown in the inset of Fig.~\ref{fig:16}. Note that the rf frequency $\omc$ is offset by the resonant frequency $\omr$ of the resonator. A sideband signal appearing in the reflection spectrum at $\omr+\omm$ is observed for sufficiently high excitation power. This measurement provides a convenient way to characterize the sensitivity of the detection method in terms of the voltage signal-to-noise ratio as the height of the sideband measured from the noise floor~\cite{ares2016prapp}. For example, for the sideband signal shown in Fig.~\ref{fig:16} we find the voltage signal-to-noise ratio as $10^{\textrm{SNR}/20}\approx 10$ measured in the bandwidth of 1~Hz. 

The above method also provides a convenient way to determine the sensitivity of the device to capacitive changes in its impedance~\cite{schoel1998Sci,ares2016prapp}. For this purpose, a voltage-tunable varactor having the capacitance $C_\textrm{v}< 1$~pF was added in parallel with the impedance of the cell (see Fig.~\ref{fig:2}(b)). By applying a sinusoidal voltage difference across the varactor, its capacitance is modulated at the frequency $\omm$ with an amplitude $\delta C_\textrm{v}= 4\times 10^{-4}$~pF, thus producing sidebands in the reflection spectrum at $\omc\pm \omm$ by the same mechanism as described earlier. From the height of the sidebands above the noise floor, the sensitivity can be determined as $S_\textrm{C}=\delta C_\textrm{v}/(\sqrt{2\Delta f}10^{\textrm{SNR}/20})$, where $\Delta f$ is the resolution bandwidth of SA~\cite{ares2016prapp}. Fig.~\ref{fig:16} shows the sensitivity $S_\textrm{C}$ determined from the height of the sidebands measured with the modulation frequency of 1.2~kHz for different values of the incident rf power $P_\textrm{in}$ measured at the output of the room-temperature source. As expected, the sensitivity increases linearly with the increasing $P_\textrm{in}$ due to the linear increase in the reflected power.

\section{Check of the electron density}
\label{app:WS}

\begin{figure}[htp]
\includegraphics[width=\columnwidth,keepaspectratio]{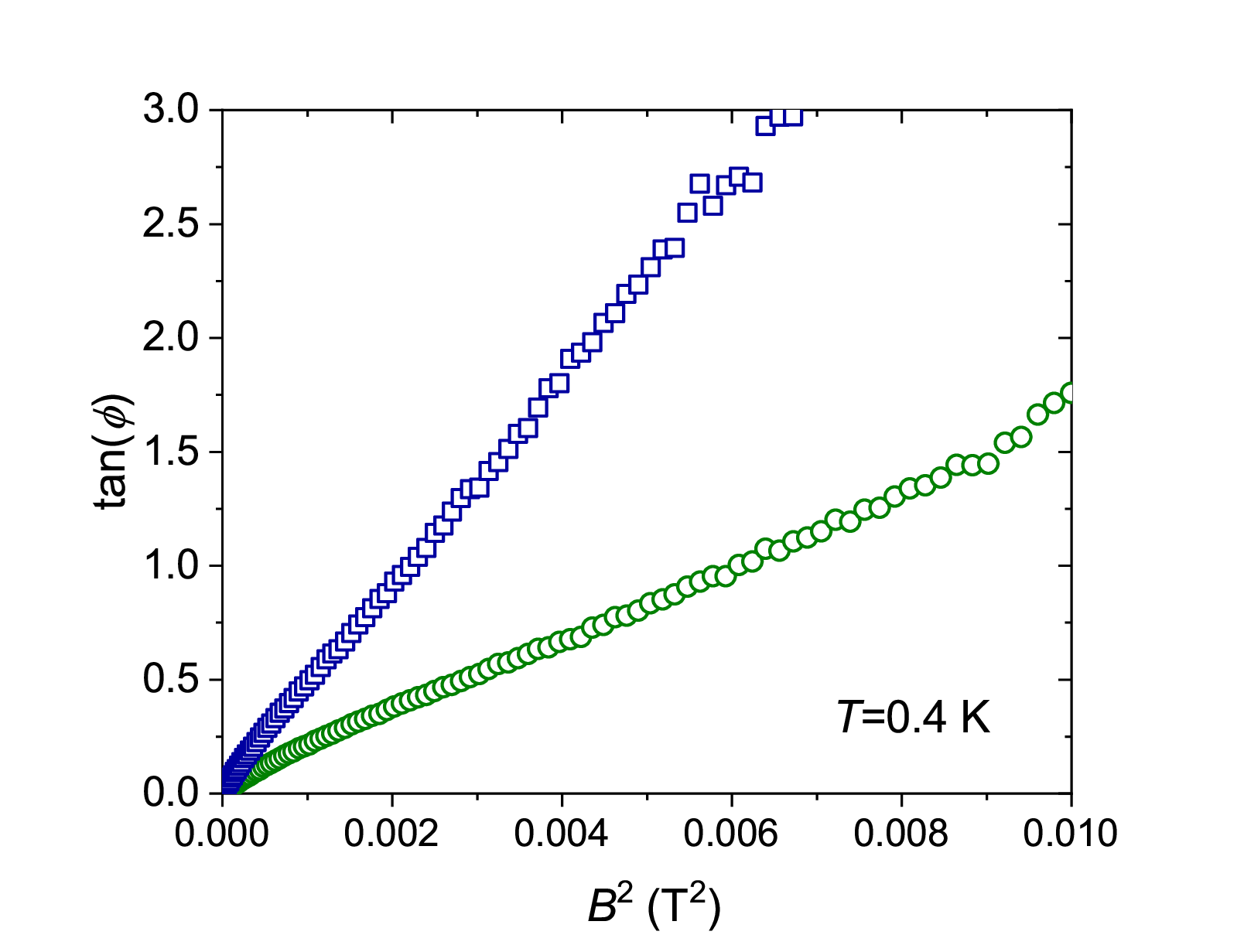}  
\caption{\label{fig:17} (color online) Tangent of the phase shift of the Sommer-Tanner signal versus the square of the magnetic field measured for electrons after the deposition and $V_\textrm{BCBM}$ set to 30~V, while all other electrodes are grounded (open circle), and after the bias voltage at the top electrode is adjusted to $V_\textrm{TCTM}=18$~V (open squares).}
\end{figure}

Usually, it is assumed that the maximum areal density of electrons accumulated on the surface corresponds to the complete screening of the electric field above the surface by the layer of the surface charge. For the surface of liquid set midway between the two capacitor plates, the corresponding density is given by $n_\textrm{s}=2\epsilon_0E_\perp/e$, where $E_\perp = (V_\textrm{BCBM}-V_\textrm{TCTM})/D$ is the pressing field exerted on the electrons (the dielectric constant of liquid helium is assumed to be unity). Nevertheless, due to the electron-electron repulsion and existence of a correlation hole of radius $r_0\sim n_\textrm{s}^{-1/2}$ surrounding each electron, the pressing field $E_\perp$ can be fully compensated by the electric field due to the surface charge $en_\textrm{s}/(2\epsilon_0)$ only at the distances from the surface larger than $r_0$. In other words, the correlation hole presents a potential barrier for an electron escaping from the surface in vertical direction, and an electron density exceeding the above estimate can be realized in this system in principle~\cite{buntarJLTP1990}. In the experimental procedure described in Sec.~\ref{sec:setup}, we start with the electron density obtained after surface charging and corresponding to the complete screening of $E_\perp$ for $V_\textrm{BCBM}=20$~V and $V_\textrm{TCTM}=0$, and subsequently decrease $E_\perp$ by setting $V_\textrm{BCBM}=30$~V and $V_\textrm{TCTM}=18$~V. Therefore, it is important to check if the electron density follows the complete screening condition which is usually assumed. For this purpose, we first carried out the conventional Sommer-Tanner (ST) measurements of the electron conductivity using the concentric Corbino electrodes comprising the bottom plate (see Fig.~\ref{fig:2}(a))~\cite{sommerPRL1971tanner}. For this measurements, the rf tank circuit was disconnected and an ac voltage with rms amplitude of 5~mV at the driving frequency of 100~kHz was applied to the center electrode of the plate, while an induced current signal was measured at the middle electrode of the bottom plate by a lock-in amplifier references to the driving frequency. This experiment was carried out in a magnetic field $B$ applied perpendicular to the surface and at the temperature $T=0.4$~K, where the electron system is in the liquid phase. The finite conductivity of the surface charge introduces a phase shift $\phi$ between the measured current signal and its purely capacitive (quadrature) component. This shift can be readily measured by the lock-in amplifier. In this experimental procedure, we first deposit electrons at $V_\textrm{BCBM}=20$~V, while all other electrodes are grounded, and then raise $V_\textrm{BCBM}$ to 30 V to stabilize the electron system. It is expected that the electron density should be about $\ns=1.1\times 10^{12}$~m$^{-2}$ from the electric-field screening after the deposition. After taking the ST measurements, we raise the top electrode voltage $V_\textrm{TCTM}$ to 18~V to set the pressing field $E_\perp=6$~kV/m, which was used in the experiments described in Sec.~\ref{sec:ex}, and repeat the ST measurements again. From the screening condition we expect to have density of about $\ns=6\times 10^{11}$~m$^{-2}$, that is about twice smaller than after the deposition. Fig.~\ref{fig:17} shows tangent of $\phi$ versus the square of the magnetic field $B^2$ for electrons after the deposition and with $V_\textrm{BCBM}$ set to 30~V (open circles), and after the bias voltage at the top electrode is adjusted to $V_\textrm{TCTM}=18$~V (open squares). For sufficiently small values of $\phi$, the inverse diagonal conductivity of the surface charge can be estimated from $\sigma_{xx}^{-1} = \tan(\phi)/(a\omega C)$, where $a$ [m$^2$] is a numerical factor determined by the dimensions of the electrodes and $C$ is the capacitance per unit area between the surface charge and the electrodes ($aC$ is on the order of pF for our setup). For sufficiently low values of the magnetic field, the dc conductivity follows the classical Drude law with $\sigma_{xx}^{-1}=(1+\mu^2B^2)/en_s\mu$, where $\mu$ is the electron mobility~\cite{leaPRB1997}. Thus, the slope of $\tan(\phi)$ versus $B^2$ dependence shown in Fig.~\ref{fig:17} is proportional to the inverse electron density $n_\textrm{s}^{-1}$. We find that the slope of the open-square plot is about twice larger than that of the open-circle plot, as should be expected from the above density estimates from the electric-field screening. Thus, this measurement confirms that upon increasing the bias voltage $V_\textrm{TCTM}$, that is decreasing the pressing field $E_\perp =(V_\textrm{BCBM}-V_\textrm{TCTM})/D$ below the screened value, the surface charge is partially lost to retain the full screening of the electric field above the surface.               

\begin{figure}[htp]
\includegraphics[width=\columnwidth,keepaspectratio]{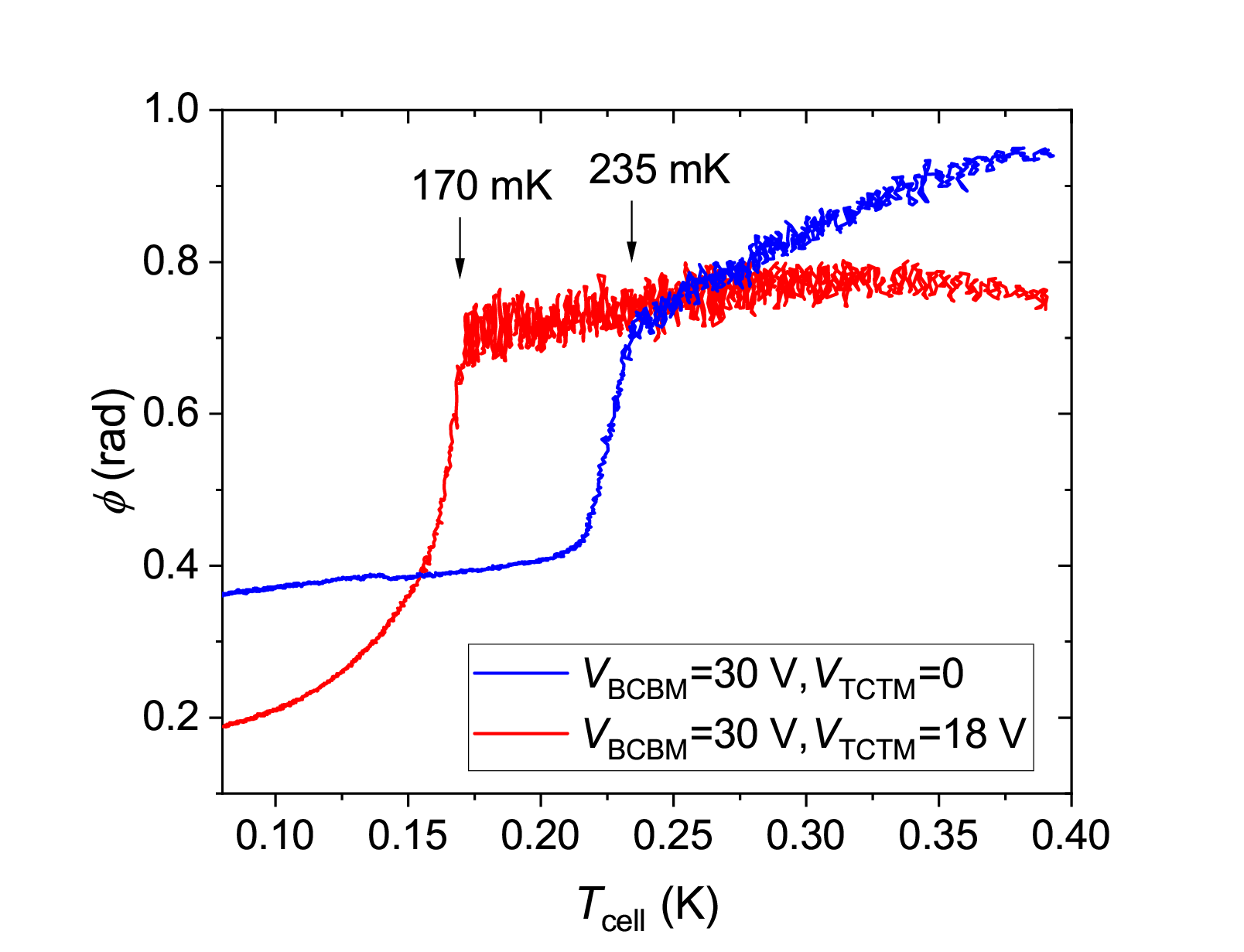}  
\caption{\label{fig:18} (color online) Phase shift $\phi$ of the ST signal recorded as a function of the temperature of the cell $T$ for electrons after the deposition and $V_\textrm{BCBM}$ set to 30~V, while all other electrodes are grounded (blue line), and after the bias voltage at the top electrode is adjusted to $V_\textrm{TCTM}=18$~V (red line).}
\end{figure}

The above method might not be sufficiently accurate to provide a reliable estimate for the absolute values of the electron density. A more reliable method to determine $\ns$ is to observe the transition of the electron system to the Wigner solid phase. The corresponding transition temperature $T_\textrm{WS}$ is related to the electron density by $T_\textrm{WS} = e^2\sqrt{\pi \ns}/(4\epsilon_0 k_\textrm{B}\Gamma_\textrm{p})$, where $\Gamma_\textrm{p}\approx 130$ is the plasma parameter corresponding to the ratio between the mean energy of the electron-electron interaction and the mean kinetic energy of electron. In this experiment, the transition of electrons to the WS phase is observed by recording the phase of the ST signal $\phi$ as a function of the temperature of the cell $T$, as shown in Figure~\ref{fig:18}. It is well established that the mobility of electrons $\mu$ drastically decreases when electrons undergo the transition into WS phase due to the coupling of electrons to the polaronic dimples, providing that the driving dc electric field is sufficiently weak. Therefore, such a transition is accompanied by a significant drop of the inverse dc conductivity $\sigma_{xx}^{-1}$, as observed in Figure~\ref{fig:18}. For the observed transition temperatures of 235~mK and 170~mK we find the electron densities $\ns =1.1\times 10^{12}$ and $5.7\times 10^{11}$~m$^{-2}$, respectively, which are in a good agreement with the values expected from the electric-field screening.

\bibliography{QEonQLSRef}

\end{document}